\documentclass[useAMS,usenatbib]{mn2e} 
\usepackage {graphicx,times}

\newcommand{\Msun}{{\rm M}_{\solar}}
\newcommand{\solar}{\ifmmode_{\mathord\odot}\;\else$_{\mathord\odot}\;$\fi}
 
\newcommand{\ma}{$K_{hb}$ } \newcommand{\mb}{$D_{hs}$ }

\begin{document}

\title[Resonances in Barred Galaxies]{Resonances in Barred Galaxies }
\author[D. Ceverino and A. Klypin]{D. Ceverino\thanks{E-mail:
ceverino@nmsu.edu} and A. Klypin\\ New Mexico State University, Las
Cruces, NM 88001}

\date{Accepted 2007 May 16. Received 2007 May 10; in original form
2007 March 28}

\pagerange{\pageref{firstpage}--\pageref{lastpage}} \pubyear{0000}

\maketitle

\label{firstpage}

\begin{abstract}
The inner parts of many spiral galaxies are dominated by bars.  These
are strong non-axisymmetric features which significantly affect orbits
of stars and dark matter particles.  One of the main effects is the
dynamical resonances between galactic material and the bar.  We detect
and characterize these resonances in N-body models of barred galaxies
by measuring angular and radial frequencies of individual orbits.  We
found narrow peaks in the distribution of orbital frequencies with
each peak corresponding to a specific resonance.  We found five
different resonances in the stellar disk and two in the dark matter.
The corotation resonance and the inner and outer Lindblad resonances
are the most populated.  The spatial distributions of particles near
resonances are wide.  For example, the inner Lindblad resonance is not
localized at a given radius.  Particles near this resonance are mainly
distributed along the bar and span a wide range of radii.  On the
other hand, particles near the corotation resonance are distributed in
two broad areas around the two stable Lagrange points. The
distribution resembles a wide ring at the corotation radius.
Resonances capture disk and halo material in near-resonant orbits.
Our analysis of orbits in both N-body simulations and in simple
analytical models indicates that resonances tend to prevent the
dynamical evolution of this trapped material.  Only if the bar evolves
as a whole, resonances drift through the phase space.  In this case
particles anchored near resonant orbits track the resonance shift and
evolve.  The criteria to ensure a correct resonant behavior discussed
in \citet{WKI} can be achieved with few millions particles because the
regions of trapped orbits near resonances are large and evolving.
\end{abstract}

\begin{keywords}
methods: N-body simulations --- galaxies: kinematics and dynamics ---
galaxies: evolution
\end{keywords}

\section{Introduction}

It is often assumed that the stellar disk in spiral galaxies can be
modeled as an axisymmetric system. Departures from this symmetry are
usually treated as weak non-axisymmetric perturbations, like spiral
arms \citep{BT87}. However, many spiral galaxies show strong
non-axisymmetric features such as central bars. In fact, galaxies with
strong bars are very common objects in the Universe.  They account for
65 per cent of bright spiral galaxies \citep{Eskr}. Barred galaxies can not
be modeled as nearly axisymmetric systems because the dynamics of
these galaxies is dominated by a strong bar which rotates around the
center. The bar interacts with galactic material and distorts galactic
orbits.  In particular, some galactic orbits experience dynamical
resonances with the bar. The motion in these orbits is coupled with
the rotation of the bar: resonant orbits are closed orbits in the
reference frame which rotates with the bar. In this frame, the bar is
stationary and a resonant orbit can periodically reach the same
position with respect to the bar. A resonant orbit is therefore a
periodic orbit in this reference frame and its dynamical frequencies
are commensurable \citep{BookDy}.

The motion of a star in a galaxy could be described by oscillations in
three dimensions: radial oscillation, an oscillation perpendicular to
the galactic plane, and an angular oscillation or rotation around the
galactic center. In general, these oscillations could be described by
three instantaneous orbital frequencies: a radial frequency $\kappa$,
a vertical frequency $\nu$ and angular frequency $\Omega$. The case of
a nearly circular orbit in an axisymmetric potential is especially
easy to understand and to study analytically using the epicycle
approximation \citep{BT87}.  However, a general orbit in the
gravitational potential of a galaxy is not a nearly circular
orbit. This is especially true for barred galaxies where the radial
oscillations are not small and orbits can be very elongated.  In
barred galaxies orbital frequencies may differ significantly from the
frequencies in the epicycle approximation.

A resonance happens if the dynamical frequencies of an orbit and the
angular frequency of the rotation of the bar, $\Omega_{B}$, satisfy
the following relationship of commensurability:
\begin{equation}
 \mathbf{l} \cdot \mathbf{\Omega} = m_B \Omega_B
\label{eq:epi}
\end{equation}
where \textbf{l}=$(l,m,n)$ is a vector of integers, $\mathbf{\Omega}$ is a
vector of frequencies, $\mathbf{\Omega}=(\kappa , \Omega, \nu )$ and
$m_B$ is also a integer. We mostly will be interested in cases with
$m_B=m$ and in motions close to the galactic plane: $n=0$. So, the
resonant condition is reduced to
\begin{equation}
 l \kappa +  m (\Omega - \Omega_{B})=0
\end{equation}
Thus, each planar resonance is described by a pair of integers,
 ($l$:$m$).  A resonant ($l$:$m$) orbit is closed after $l$
 revolutions around the center and $m$ radial oscillations in the
 reference frame which rotates with the bar. Table 1 summarizes
 several examples of these resonances.

One important example is the corotation resonance (CR), $ \Omega=
\Omega_{B}$. A star in a resonant orbit of CR rotates around the
galaxy center with a speed equal to the rotation speed of the
bar. Thus, it does not move in the rotating frame. Other important
resonances are the inner and outer Lindblad resonances. As it seems
from the rotating frame, a star in one of these resonant orbits makes
two radial oscillations during one angular revolution. The resonant
orbit has therefore an ellipsoidal shape. The inner Lindblad resonance
(ILR) corresponds to the (-1,2) resonance. In this case, the star
rotates faster than the bar, $\Omega > \Omega_{B} $. The opposite
case, $\Omega < \Omega_{B} $, corresponds to the (1:2) resonance,
which is the outer Lindblad resonance (OLR).

\begin{table} 
\caption{Examples of resonances}
 \begin{center}       
 \begin{tabular}{|lllcccc} \hline     
\multicolumn{2}{l} {Name} $l$ & $m$ & $n$ &
$\frac{\Omega-\Omega_B}{\kappa}$ \\\hline CR & 0 & 1 & 0 & 0 \\ ILR &
-1 & 2 & 0 & 0.5 \\ OLR & 1 & 2 & 0 & -0.5 \\ UHR & -1 & 4 & 0 & -0.25
\\
 \end{tabular} 
 \end{center}
 \end{table}

Significant effort have been made to study motions near resonances in
astrophysical systems as well as in plasma physics and in Solar system
dynamics \citep{Chirikov,LBK72,TW84,W04,WKI}.  One important and
difficult problem is the problem of small divisors or small
denominators. Suppose we impose a small perturbation and study the
effect of this perturbation. In the case of barred galaxies, the
unperturbed case is an axisymmetric model and the perturbation is a
weak bar.  One may try to find a solution to this problem using
perturbation series.  When this is done, the solution typically has
terms with denominator $ \mathbf{l} \cdot \mathbf{\Omega} -m_B
\Omega_B$, which goes to zero at resonance.  The reason for this
divergence is the breakdown of the assumption that the solution can be
written as a perturbation series. It appears that correct behavior of
the solution cannot be obtained in any order of the perturbation
theory \citep[Sec.2.2b]{BookDy}. One may think that the perturbation
theory gives qualitatively correct answer (e.g., predicts large
changes in energy and angular momentum), but it fails to estimate the
magnitude of the effect. Unfortunately, this is not the case: it gives
wrong qualitative answers.  \citet[Chapter 3,
eqs. (3-123)-(3-129) ]{BT87} give an example of treatment of orbits
around stable corotation resonance (Lagrange points $L_4$ and
$L_5$). In this case the perturbation expansion gives a divergent
amplitude (the solution linearly grows) and the correct treatment in
\citet{BT87} does not show {\it any} growth (see also
\citet{BFButa06}).

It is important to formulate the situation clearly because this can
produce significant confusion. In mathematics the problem is often
stated as the problem of perturbations: how orbits change when a
perturbation is imposed. In this case one compares perturbed orbits
with the same orbits before the perturbation was imposed. Significant
deviations are expected to happen for unperturbed trajectories in
regions of overlapping resonances of the unperturbed system
\citep{Chirikov}. Yet, this is not the problem, which we deal with in
barred galaxies. In this case we study only perturbed orbits: how they
change with time and how they behave close to resonances of the
perturbed system. In other words, we do not compare perturbed orbits
with the unperturbed trajectories. By itself it is a very interesting
problem: formation of bars. Yet, at this moment we focus exclusively on
the evolution under the forces of bars.

This is significantly easier problem. We also simplify the situation
by considering bars which do not change with time. First, we start
with orbits at exact resonances: $\vec{l} \cdot \vec{\Omega} = m_B
\Omega_B$. How do they evolve?  The answer is simple: they do not
\citep{LBK72}. Orbits at exact resonances are closed in the
phase-space: after some time they come to exactly the same position in
space and have exactly the same velocities. Thus, they have the same
angular momentum and the same energy. 

Close to the resonances the situation is complex. \citet{LBK72} argued
that there should be significant growth of perturbations in this area.
Yet, this argument was based on the perturbation expansion, which is
not valid near resonances. We distinguish two types of resonances:
elliptic and hyperbolic \citep{Arnold}. Lagrange points $L_4$ and
$L_5$ are examples of elliptic resonances: orbits oscillate and
librate around those resonances and have the structure of a simple
pendulum 
(Lichtenberg \& Lieberman 1983, Sec. 2.4, Murray \& Dermott 1999, Sec. 8). 
Hyperbolic
resonances are points on intersection of separatrixes dividing domains
of elliptical resonances (e.g., Lagrange points $L_1$ and $L_2$)
In this paper we are mostly interested in elliptical resonances. 

There is no evolution at resonant orbits for a stationary perturbation
and there is no singularities at resonances. Therefore, the small
divisor problem can lead to a wrong interpretation of the secular
evolution near resonances and their role in barred galaxies.  A more
careful treatment of the motions in near-resonant orbits reveals that
the Hamiltonian near a resonance can be approximated by the
Hamiltonian of the one dimensional pendulum in a variable which change
slowly near the resonance \citep{BookDy}. So, the motions near every
resonance can be approximated by motions of libration, separatrix and
rotation around a resonant orbit. Examples of these motions near CR
and ILR can be found in section 6. Circulating pendulum solutions near
CR are also given by \citet{BFButa06}.  In general, each fixed point
of the pendulum corresponds to an exactly resonant orbit. As the
result, each resonance has formally two different types of resonant
orbits. One corresponds to the stable or elliptic fixed point, around
which the near-resonant orbits librate. The other corresponds to the
unstable or hyperbolic fixed point, where the separatrices
intersect. In the region around the separatrix, this pendulum
approximation fails. The phase-space near the separatrix is more
complex that in the case of a pendulum \citep{Athens}. This area may
be filled with irregular or chaotic orbits and high-order resonances
($l>1$). This is usually called the resonance layer \citep{BookDy}. As
a result of all this complexity, perturbation theory is not valid at
any order near resonances. Small divisors can be removed from one
order in the perturbation series, but other small divisors appear in a
higher order. So, the right behavior near resonances can only be
studied by solving the exact solution of the equations of motions near
resonances.

The phase-space near resonances is mainly populated by trapped orbits
in libration around stable resonant orbits. Their exact trajectories
are commonly computed in orbit theory \citep{Contop89,Skokos}. In
this field, the potential of a barred galaxy is modeled by a
combination of different analytical potentials, like an axisymmetric
disk plus a prolate ellipsoid. Then, galactic orbits are computed
numerically using this galactic model. In this way, the galactic
orbital structure can be studied in detail. Resonant orbits in this
case are periodic orbits in a given non-axisymmetric potential. Each
stable periodic orbit is the parent of a family of non-closed orbits
which remain near to this orbit at any moment \citep{BT87}. The
dynamical frequencies of these trapped orbits oscillate around the
frequencies of the resonant orbit. Therefore, their average
frequencies over time should be close to the frequencies of the
resonant or parent orbit.

However, orbit studies have some limitations. They can not follow the
self-consistent evolution of barred galaxies.  The underlying
potential is fixed and does not change due to the redistribution of
the orbits. In contrast, N-body models can follow the orbits and the
secular evolution of barred galaxies at the same time. However,
\citet{WKI} have derived the necessary number of particles in an
N-body model which could accurately resolve the dynamics near
resonances. These required numbers are well beyond the numbers used in
current state-of-the-art models. So, do we have any hope to see the
effects of resonances in N-body models? We argue that current N-body
models can resolve the dynamics of resonances in the regime relevant
for observed barred galaxies. They have strong non-axisymmetric
features. In contrast, the particle number criteria of Weinberg and
Katz (2007a) were derived in the regime of weak perturbations.

N-body models have been already used to study the resonant interaction
between the bar and the halo of dark matter
\citep{HB05,CVK06,Atha02,Atha03,Martinez,WKII}. N-body models open the
possibility to sample individual trajectories over time and extract
their dynamical frequencies. This allows a better determination of
resonant orbits and their dynamics. This has been done in restricted
N-body experiments with a frozen non-axisymmetric potential
\citep{HB01,Atha02,Atha03,Martinez}. In \citet{Atha02}, the orbital
frequencies were estimated using a random population of particles
taken from the disk and the halo of a N-body simulation. A frozen
barred potential equal to the potential of the simulation was set to
rotate with the pattern speed measured in the simulation at a given
moment. Each orbit was computed in this stationary potential. Finally,
the dynamical frequencies of each orbit were estimated using a
spectral analysis. Some of the orbits were trapped near resonant
orbits in the disk and in the halo. The slowdown of the bar was linked
to the lost of angular momentum of nearly resonant orbits in the inner
disk. At the same time, the gain of angular momentum of the halo was
linked to the gain of angular momentum of near-resonant orbits in the
halo.

However, little work has been published on the detection of resonances
in a fully self-consistent N-body model of a barred galaxy. The
purpose of this study was to detect and characterize the resonances
present in barred galaxies. This study may also find new insights into
the dynamics near resonances and their role in barred galaxies.  This
paper is organized as follows. \S2 presents the N-body models analyzed
in this paper. \S3 describes the methods used to measure the dynamical
frequencies of the particles. \S4 describes the main results on
resonances in the disk and in the halo. \S5 describes the capture at
corotation as an example of resonant capture. \S6 compares these
results with an analytical galactic model. Finally, \S7 is devoted to
the discussion and \S8 is the summary and conclusion.

\begin{figure*}
\includegraphics[width =\textwidth]{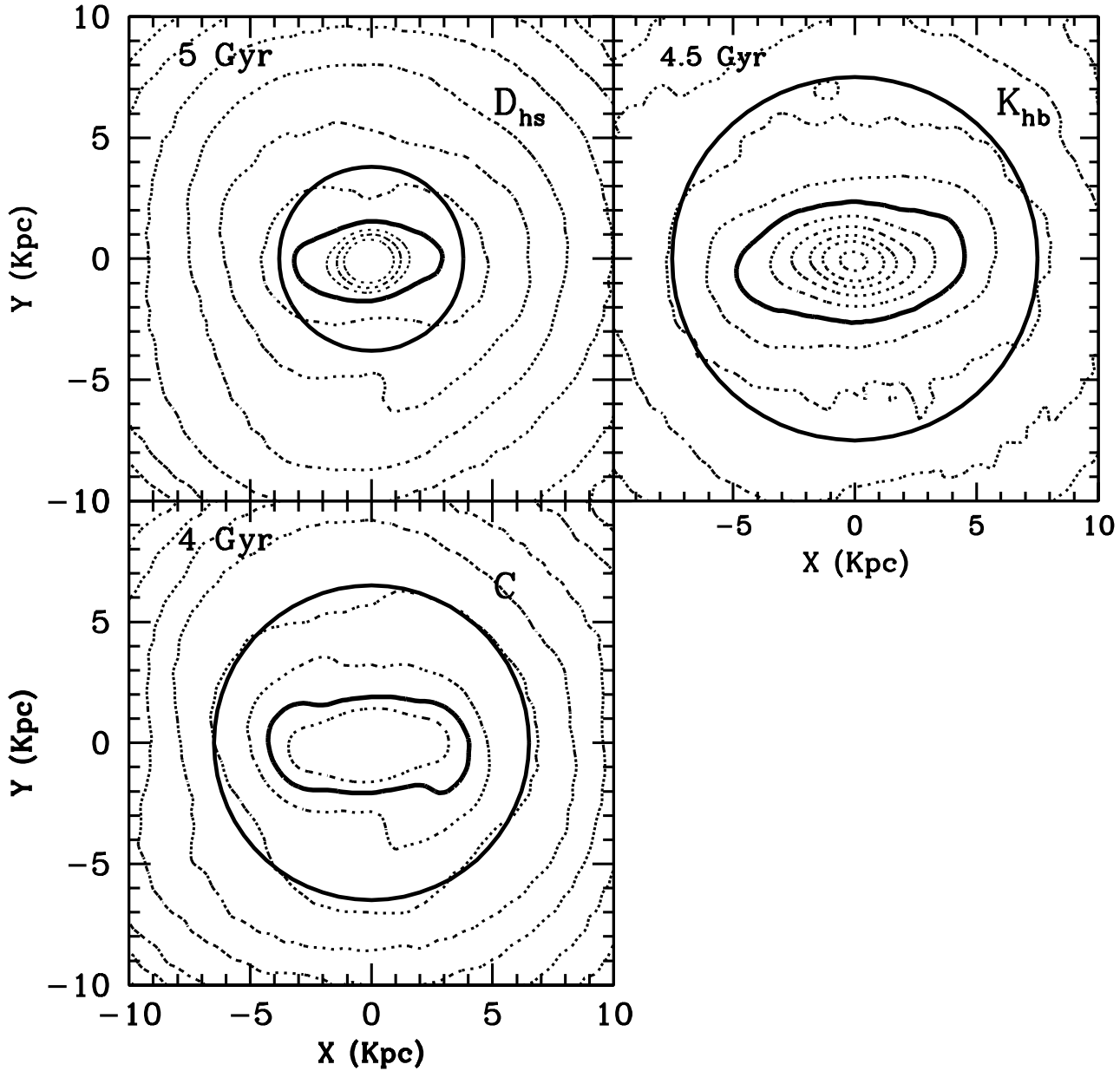}
 \caption{Face-on views of the three N-body models of barred galaxies.
The contours show levels of equal surface density.  Their scale is
logarithmic with a step of 0.5 dex.  The scale is the same for all
figures.  The circles represent the corotation radius for each model.
The model $K_{hb}$ develops a larger and slower bar than the model
$D_{hs}$.  The model C has a strong and massive bar.}
\label{fig:1}
\end{figure*}

\section{The models}
\begin{table} 
\caption{Initial parameters of models}
 \begin{center}       
 \begin{tabular}{|lllcccc} \hline     
\multicolumn{2}{l} {Parameter} $ D_{hs}$ & $K_{hb}$ & C \\\hline Disk
 Mass ($10^{10} \Msun$) & 5.0 & 5.0 & 4.8 \\ Total Mass ($10^{12}
 \Msun$) & 1.43 & 1.43 &1.0 \\ Disk exponential length (kpc) & 2.57 &
 3.86 & 2.9 \\ Disk exponential height (kpc) & 0.20 & 0.20 & 0.14 \\
 Stability parameter $Q$ & 1.8 & 1.8 &1.2 \\ Halo concentration C & 17
 & 10 & 19 \\ Total number of particles ($10^5$)& $38.0$ & $27.2$ & $
 97.7$ \\ Number of disk particles ($ 10^5$) & $4.60$ & $2.33$ & $
 12.9 $ \\ Particle mass ($10^5 \Msun$) & $1.07$ & $2.14$ & $0.37$ \\
 Maximum resolution (pc) & 22 & 22 & 100.\\ Time Step ($10^4 yrs$) &
 1.48 & 0.95 & 12. \\
 \end{tabular} 
 \end{center}
 \end{table}

\subsection{Initial conditions}

The initial conditions of the N-body models are described in detail in
\citet{VK03}. The generation of the models follows the method of
\citet{Hernquist93}. The model initially has only a stellar
exponential disk and a dark matter halo. No bar is initially present
in the model but the system is unstable and forms a bar.  
The density of the stellar disk in cylindrical
coordinates is approximated by the following expression:

\begin{eqnarray}
 \rho_d(R,z) =\frac{M_d}{4\pi z_0R_d^2}e^{-\frac{R}{R_d}}
                       sech^2(z/z_0),
\end{eqnarray}
where $z_0$ is the scale height of the disk, $R_d$ is the exponential
length and $M_d$ is the mass of the disk.  The scale height is assumed
to be initially constant through the disk.  The vertical velocity
dispersion $\sigma_z$ is given by the scale height and the surface
stellar density $\Sigma$:
\begin{eqnarray}
 \sigma^2_z(R) =\pi G z_0 \Sigma(R),
\end{eqnarray}
where G is the gravitational constant.  The radial velocity dispersion
$\sigma_R$ is also related to the surface density:
\begin{eqnarray}
  \sigma_R(R) =Q\frac{3.36G\Sigma(R)}{\kappa(R)},
\end{eqnarray}
where $\kappa(R)$ is the epicycle frequency at a given radius and Q is
the Toomre stability parameter, assumed constant through the disk.
The rotational velocity $V_\phi$ and its dispersion $\sigma_\phi$ are
computed using the asymmetric drift approximation and the epicycle
approximation,
\begin{eqnarray}
 V_\phi^2(R) &=& V_c^2(R) -\sigma_R^2(R)\left(\frac{2R}{R_d}+
                  \frac{\kappa^2(R)}{4\Omega^2(R)}-1\right), \\
\sigma_\phi^2(R) &=& \sigma^2_R(R) \frac{\kappa^2(R)}{4\Omega^2(R)}, 
\end{eqnarray}
 where $V_c$ is the circular velocity at a given radius and $\Omega$
 is the angular frequency in the epicycle approximation.

The density profile of a cosmological motivated dark matter halo is
initially well approximated by the NFW profile \citep{NFW},
\begin{eqnarray}
 \rho_{\rm dm}(r) &=& \frac{\rho_s}{x(1+x)^2}, \ x=r/r_s, \\ M_{\rm
 vir} &=& 4\pi\rho_s r_s^3\left[ \ln(1+C) -{C\over 1+C}\right], \
 C={r_{\rm vir}\over r_s},
\end{eqnarray} 
where $ M_{\rm vir}$ and C are the virial mass and the concentration
of the halo.  The radial velocity dispersion of dark matter particles
is related with the mass profile of the system, M(R),
\begin{eqnarray}
  \sigma_{R, {\rm dm}}^2 ={1\over \rho_{\rm dm}}\int_R^{\infty}\rho_{\rm dm}\frac{GM(R)}{R^2}dR.
\end{eqnarray}
Finally, the other two components of the velocity dispersion of dark
matter are equal to $\sigma_{r, {\rm dm}}^2$, assuming an isotropic
velocity distribution. This assumption remains valid in the central
parts of dark matter halos \citep{Colin00}.

\subsection{Description of the models}

We analyze two of the N-body models of barred galaxies described in
\citet{CVK06}. We also include the model C of \citet{VK03}. These
three models are consistent with normal high surface brightness
galaxies. The dark matter does not dominate in the models in the
first two scale lengths, $R\leq2R_d$ \citep{Klypin02}. The parameters
of the models are presented in Table 2. These models do not cover a
large range of parameters. This is done in \citet{CVK06}. Instead, we
have selected three models with very different initial conditions. \mb
has a more concentrated halo and a shorter disk length than
$K_{hb}$. For example, the dark matter contribution to the initial
circular velocity is equal to the contribution of the disk at 7 Kpc in
the model \mb and 10 Kpc in the model $K_{hb}$. As a result, \mb is
initially more centrally concentrated than $K_{hb}$. The disk is hot,
$Q\approx 1.8$, in these two models. In contrast, the model C has a
cold disk, $Q\approx 1.2$. In addition, the halo of this model has a
higher concentration than the models from \citet{CVK06} but its
exponential length is in between the values of the other two
models. As a result, 6.5 Kpc is the radius in which the contribution
of the halo and the disk to the initial circular velocity are
equal. All these differences are reflected in the bar evolution and
therefore, they are also reflected in the resonant structure.  All
three models develop a relatively strong bar (Fig.
\ref{fig:1}). However, the bar in the model \mb is shorter and rotates
faster than the bar in the model $K_{hb}$. This affects strongly the
resonant structure.  All three models show a slow evolution in the
pattern speed of the bar, $\Omega_B$, so it could be considered nearly
constant over a period of 1-2 Gyr \citep{CVK06}. This is a suitable
situation for the analysis of resonances because the resonant
structure is stable if $\Omega_B$ is constant.
\subsection{The code}

These simulations were performed with the Adaptive Refinement Tree
(ART) N-body code \citep{Kravtsov97,Kravtsov99}. The code computes
the density and gravitational potential in each cell of a uniform
grid. If the number of particles in a cell exceeds a given threshold,
the cell is split in 8 smaller cells. This creates the next level of a
refinement mesh. The procedure is recursive. The result is a refined
mesh which accurately matches high density regions with arbitrary
geometry.  This spatial refinement is followed by a temporal
refinement. More refined regions have a shorter time step. This is
necessary to follow accurately the trajectory of particles. The code
was extensively tested. Additional tests on the long-term stability of
equilibrium systems were performed in \citet{VK03}.  These tests are
important to study the secular evolution in barred galaxies.  The
results showed that the effect of two-body scattering is
negligible. The relaxation time scale was roughly equal to 
$4.5\times 10^4$ Gyr for a system with 3.5 million particles.

\begin{figure}
\includegraphics[width =0.45\textwidth]{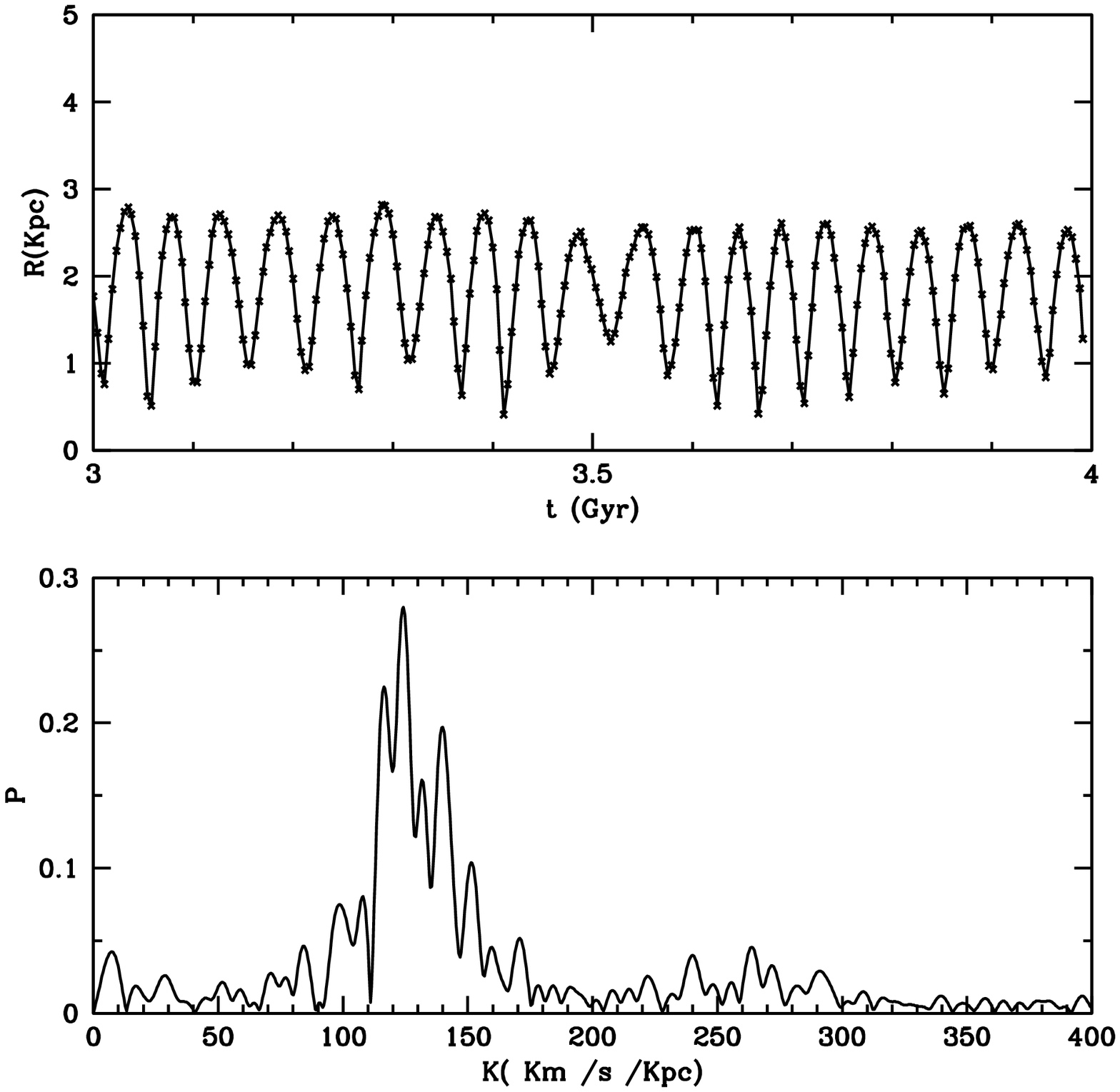}
 \caption{Example of a measurement of a radial frequency, $\kappa$, of
a particle using the Fourier spectrum of its radial oscillations. The
top panel shows the radius of a particle as a function of time for \ma
and the bottom panel presents the Fourier decomposition of its radial
trajectory, where P is the power spectrum.  The radial frequency is
measured as the maximum peak of the spectrum, $\kappa=126$ Km s$^{-1}$
Kpc$^{-1}$.}
\label{fig:2}
\includegraphics[width =0.45\textwidth]{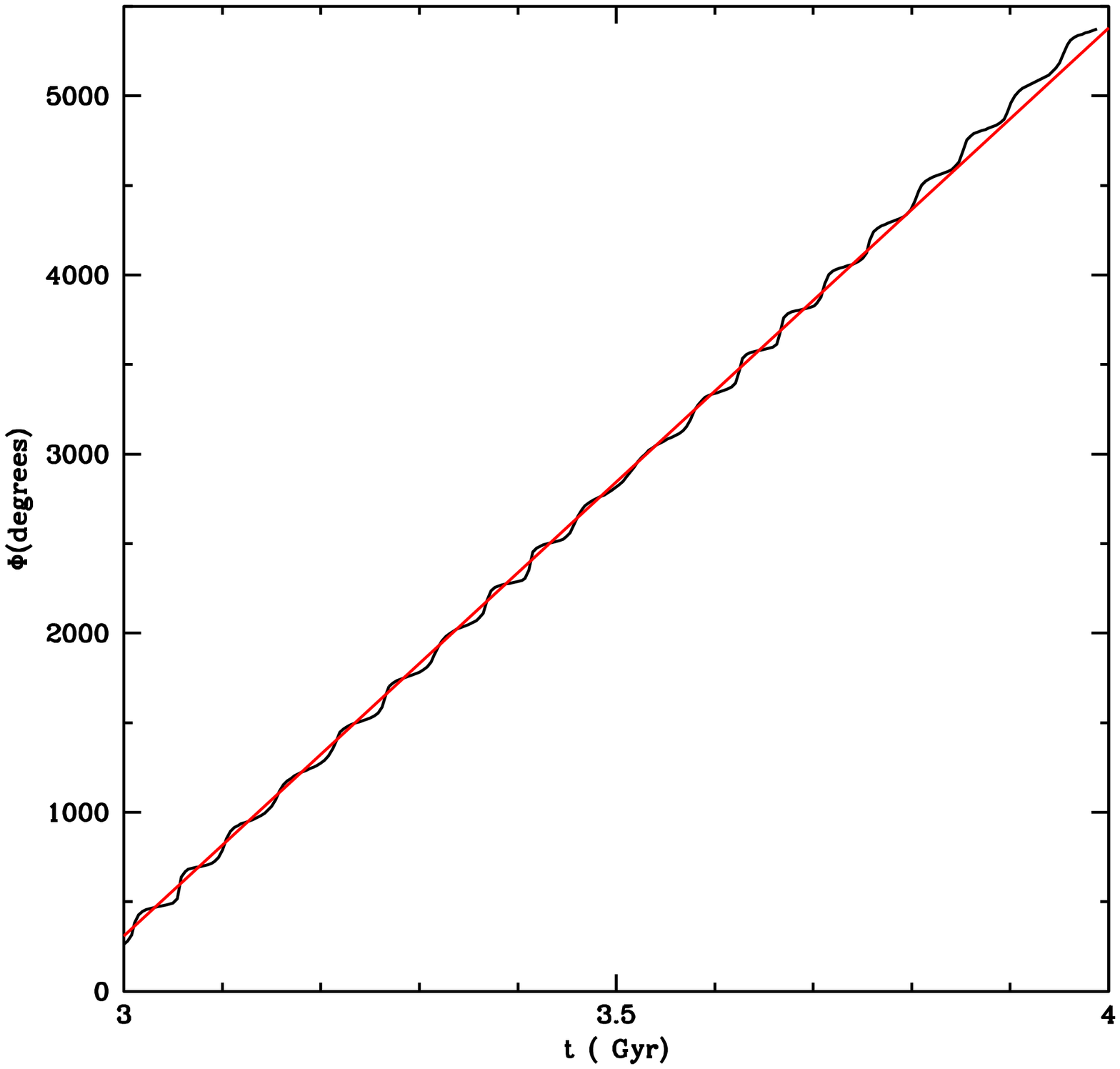}
 \caption{Example of a angular frequency measurement of a particle
using its angular position along its trajectory.  The Figure shows the
angular positions over time for a particle selected in $K_{hb}$.  The
straight line has a slope equal to the measured angular frequency,
$\Omega = 88$ Km s$^{-1}$ Kpc$^{-1}$.}
\label{fig:3}
\end{figure}
\section{Measurement of orbital frequencies}
 
We measure the orbital frequencies of all particles over a given
period of time. This time average is an estimate of the instantaneous
orbital frequencies used in the resonant condition (Eq. 2). The
measurements are done using the trajectories of all particles. In the
models \ma and $D_{hs}$, each trajectory is sampled with $N_{t}=250$
discrete points per Gyr. Therefore, two consecutive points are
separated by $ 4 \times 10^6$ yr.  Each trajectory during a single
sampling step is integrated with more than 270 time steps.  The
trajectories are recorded in cylindrical coordinates. The orbital
frequencies are estimated by tracking the radius and the azimuthal
angle as functions of time.

The radial frequency, $\kappa$, is measured from the Fourier analysis
of the radial oscillations.  Fig. \ref{fig:2} shows an example of
the radial oscillations for one trajectory.  We subtract the average
radius from the signal and perform a discrete Fourier analysis.  The
result is the power spectrum of the orbit $(P_{k})$.  It is based on
the harmonics amplitudes, $A_{k}$ and $B_{k}$, in the Fourier
decomposition:
\begin{equation}
A_{k} =\frac{1}{N_{t}} \sum_{i} R_{i}\cos(\omega_{k}t_{i})  , 
B_{k} =\frac{1}{N_{t}} \sum_{i} R_{i}\sin(\omega_{k}t_{i})  
\end{equation}
\begin{equation}
P_{k} = \sqrt{ A_{k}^{2} +  B_{k}^{2}}.
\end{equation}
where $ R_{i}$ is the value of the radius at a given time $t_{i}$ and
$\omega_{k}$ is a discrete frequency in the Fourier space. We use
$1200$ discrete frequencies to sample the Fourier space from zero to a
maximum frequency of 600 Km s$^{-1}$ Kpc$^{-1}$ for $D_{hs}$.  This
maximum frequency is well below the Nyquist frequency.  In our case,
the Nyquist frequency $( \kappa_{N} = \pi / \Delta ) $ is 770 Km
s$^{-1}$ Kpc$^{-1}$, where $\Delta$ is the interval of $4 \times 10^6$
yr between snapshots.  In that way, we avoid aliasing problems that
arise close to the Nyquist frequency.  The bottom panel of the Fig.
\ref{fig:2} shows an example of the spectrum of the trajectory.  The
radial frequency is measured as the frequency of the maximum peak in
the Fourier spectrum.

However, the gravitational potential is slowly evolving during the
period in which the frequencies are measured. This introduces radial
modes of low frequency at the top of the orbital oscillations.  As a
result, 5 per cent of the particles have radial oscillations modulated by
an oscillation of low frequency.  For these particles, we need to
remove these low frequency modes to be able to extract the orbital
oscillations.  In order to remove these modes from the signal, we
define a low cutoff frequency of 12 Km s$^{-1}$ Kpc$^{-1}$.  If the
frequency of the maximum peak of the spectrum is bellow this cutoff,
the corresponding mode is subtracted from the radial
oscillation. Then, we repeat the Fourier analysis.  The procedure ends
when the maximum of the spectrum lies beyond the cutoff frequency or
when we remove all the significant peaks of the spectrum.  In the last
case, we reject the particle because its trajectory does not have
significant radial oscillations.  However, this technique prevents us
to detect radial frequencies lower than the cutoff frequency.  These
radial oscillations would correspond to trajectories in the edge of
the disk, where the effect of the bar is very small. So, these
trajectories are not useful for study resonances.

The spectral analysis used for radial frequencies was proved less
reliable for angular frequencies \citep{Atha02}. In contrast with
radial oscillations, the azimuthal angle does not oscillate around a
mean value. The angle sweeps periodically all values between 0 and
$2\pi$. As a result, the angular frequency is measured using the
average period of the angular revolutions in that interval of time.
Each angular period is defined as the time that the particle takes to
complete one angular revolution starting from a given point of the
trajectory. Fig. \ref{fig:3} shows an example of the angular
positions of one trajectory and the computed angular frequency.

We performed an orbital frequency analysis of all particles in the
three models. The following orbits were rejected from a further study:
Retrograde orbits, orbits with radial frequencies higher than a
maximum frequency of 600 Km s$^{-1}$ Kpc$^{-1}$ and orbits with none
significant radial oscillations $(\kappa \le 12$ Km s$^{-1}$
Kpc$^{-1}$). In total, we rejected only 10 per cent of the particles in \ma
and 30 per cent of the particles in $D_{hs}$.  The higher fraction in \mb
is due to a higher concentration of particles at the center. One half
of the rejected particles in \mb have very high frequencies and almost
radial orbits. They expend all the time very close to the center. So,
they are not involved in global motions with the bar. As a result, we
selected only particles with well defined orbital frequencies during a
given interval of time.

\section{Results}
\begin{figure}
  \includegraphics[width =0.45\textwidth]{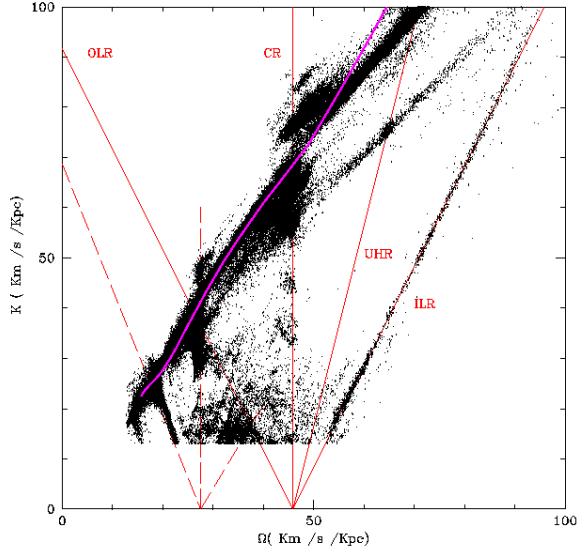}
 \caption{Frequency map for $D_{hs}$. Horizontal axis shows angular
frequencies $(\Omega)$ and vertical axis shows radial frequencies
$(\kappa)$. Each point represents the average frequencies of a
particle over 1 Gyr.  Clustering of points along straight lines with
certain slopes indicates the presence of resonances.  The lines are
computed using the resonant condition (eq. 2).  $\Omega_B$ is computed
as the average pattern speed in this period of time. The main
resonances are labeled.  In addition, the dash lines correspond the
CR, ILR and OLR of another non-axisymmetric pattern with a different
pattern speed.  The grey curve (magenta in the color version) is the
result of the epicycle approximation which breaks inside the
corotation radius, where the orbits are very elongated }
\label{fig:4}
\end{figure}
\begin{figure}
\includegraphics[width =0.45\textwidth]{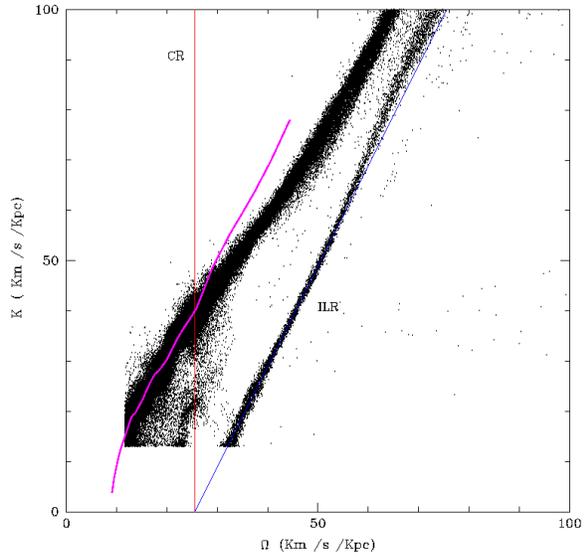}
 \caption{Frequency map for \ma (massive bar) for a period of 0.5 Gyr.
As Fig. \ref{fig:4}, lines indicate the main resonances and the grey
curve (magenta in the color version) represents the epicycle
approximation.  The resonant structure shows a strong clustering of
points along the ILR line. This is an indication of a large number of
particles trapped near ILR orbits.}
\label{fig:5}
\end{figure}  
\subsection{Detection of the main resonances}

Once the orbital frequencies are measured in our three models, their
main resonances become evident using frequency maps. They are commonly
used to study resonances between planetary orbits in the Solar system
\citep{Laskar} and in orbital studies of elliptical galaxies
\citep{HB01}.  In our case, a frequency map displays angular
frequencies along the horizontal axis and radial frequencies in the
vertical axis. This is a clear way to display the resonant structure
of a model in the space of its orbital frequencies. Each point in this
space represents the mean orbital frequencies of an individual
particle over a fixed period of time. In a frequency map, all the
orbits near a particular resonance lie along a line defined by the
resonant condition (Eq. 2). $\Omega_B$ is computed as the average
pattern speed in this period of time \citep{CVK06}. Any set of
integers defines a line in the frequency map. As a result, points
along a given resonant line correspond to particles close to the
resonance by definition.
\begin{figure}
\includegraphics[width =0.45\textwidth]{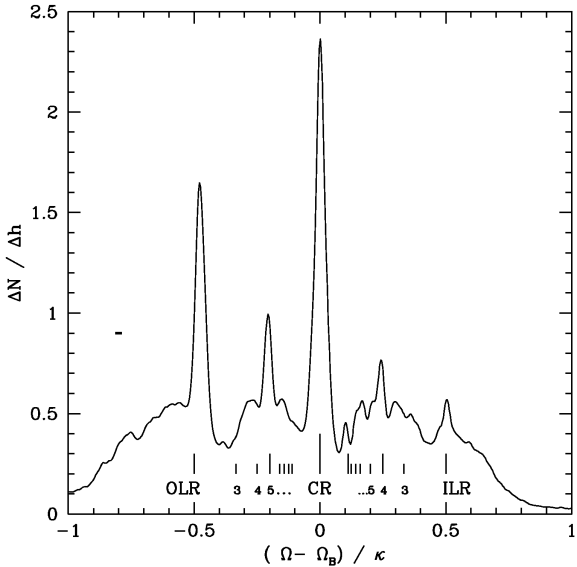}
 \caption{Distribution of the ratio $ ( \Omega - \Omega_{B} ) /
\kappa$ for \mb for a period of 1 Gyr.  The vertical axis shows the
fraction of particles per unit of bin in the frequency ratio.
Vertical lines represent low order resonances ($\pm1$:m) and CR.  The
peaks show a strong indication of trapping resonances.  The error-bar
at the upper-left corner is the $1\sigma$ error using Poisson noise.}
\label{fig:6}
\end{figure} 
\begin{figure}
\includegraphics[width =0.45\textwidth]{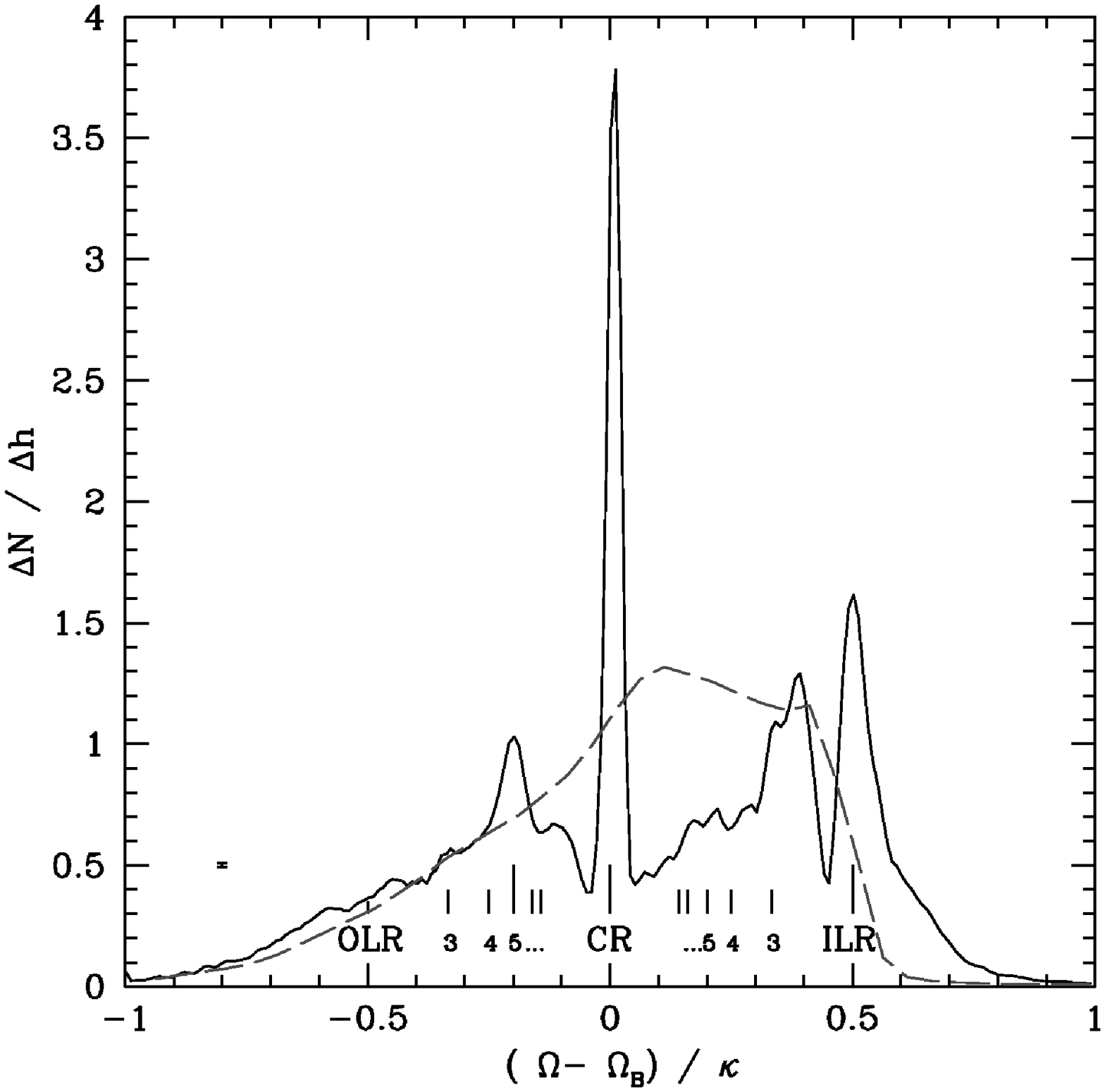}
 \caption{Distribution of the ratio $ ( \Omega - \Omega_{B} ) /
\kappa$ for \ma for a period of 2 Gyr.  The vertical axis shows the
fraction of particles per unit bin in the frequency ratio.  The
error-bar on the left is the $1\sigma$ error using Poisson noise.
Strong peaks at CR and ILR are clearly present.  The dash line shows
the distribution for a period of 0.5 Gyr before the formation of the
bar.  No peaks are found before the formation of the bar. The
formation of these resonant peaks is linked to the capture of
particles near resonant orbits.}
\label{fig:7}
\includegraphics[width =0.45\textwidth]{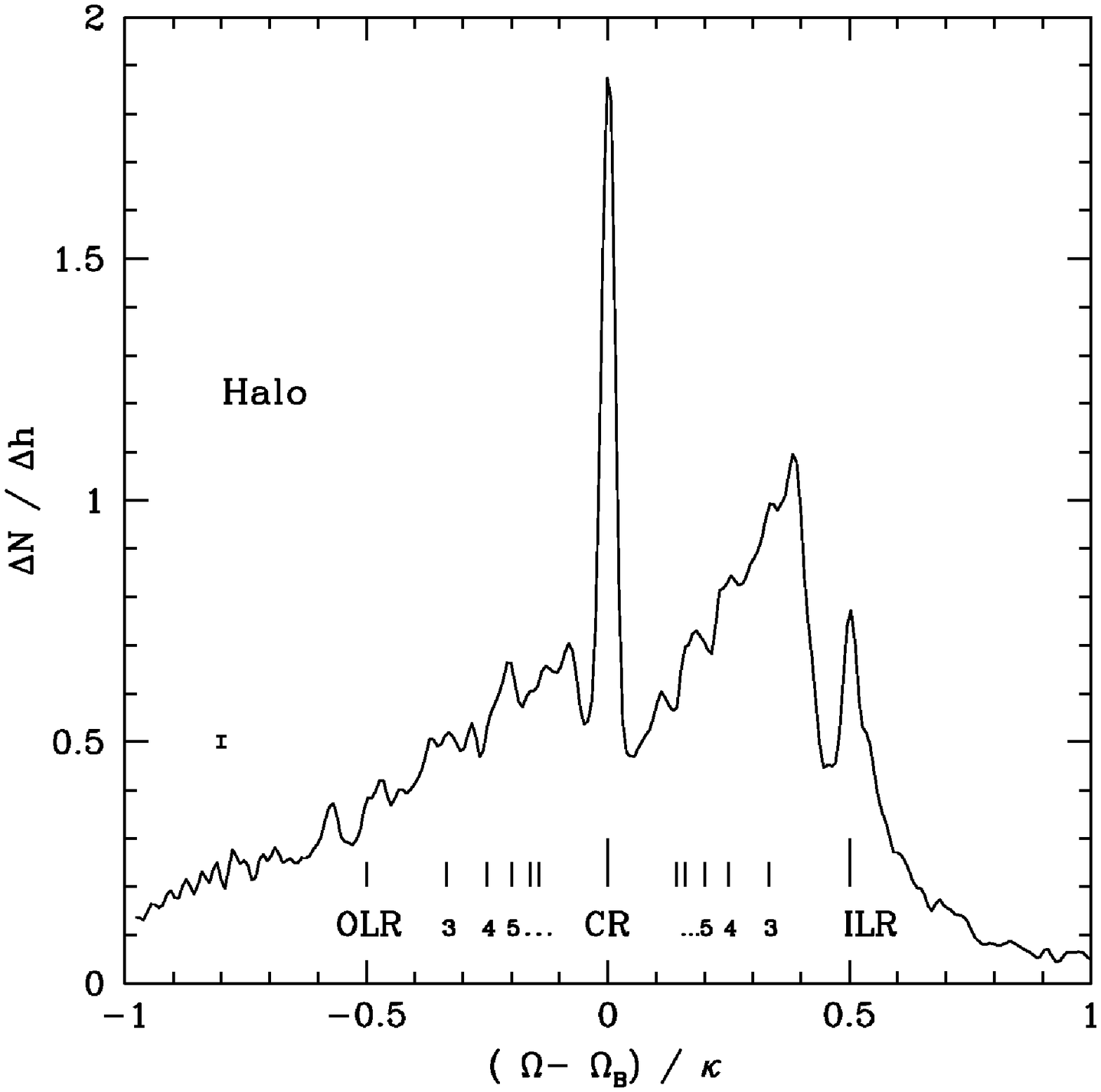}
\caption{Distribution of the ratio $ ( \Omega - \Omega_{B} ) / \kappa$
for particles in the halo chosen to stay close to the disk of
$K_{hb}$. The lines shows resonances as Fig. 6 and 7.  The CR and ILR
are clearly present in the halo.  The error-bar is the $1\sigma$ error
using Poisson noise.}
\label{fig:8}
\end{figure}

Fig. \ref{fig:4} shows the frequency map of the model $D_{hs}$ for 1
Gyr. We can clearly see concentration of particles near resonances. In
particular, a narrow line of points is clearly visible along the
ILR. This resonance covers a big range of angular and radial
frequencies. The concentration of orbits near the CR is also
especially strong. These particles are forced by the bar to rotate
with the bar pattern speed.  Other resonances are also visible. For
example, the ultra-harmonic (-1:4) resonance (UHR) can be seen as a
small cloud of points which intersects the line corresponding to the
UHR.  The OLR is very prominent. It intersects a line of constant
angular frequency, $\Omega=27$ Km s$^{-1}$ Kpc$^{-1}$.  This may be
the CR of another non-axisymmetric pattern. This is supported by the
fact that both OLR and ILR are also present for this second
pattern. The OLR line is specially well populated. As a result, the
model \mb may exhibit an overlap of resonances corresponding to two
different non-axisymmetric features.

As another example, Fig. \ref{fig:5} shows the frequency map for the
model $K_{hb}$ for a period of 0.5 Gyr.  This model has a more massive
bar which also rotates slower than in the model $D_{hs}$ discussed
before. As a result, the resonant structure appears more compressed in
this model. Its points extend over a smaller fraction of the
frequency space.  In addition, the corotation radius is larger.  So,
resonances beyond corotation, $\Omega<\Omega_B$, extend over the outer
disk and they have less available material to capture.  Therefore, OLR
is not even present in this model.  On the other hand, ILR is much
stronger. It has more points clustered along the ILR line. This is the
effect of a massive bar.  The clustering of points near resonant lines
in the frequency maps is a signature of trapped particles near
resonant orbits.
\begin{figure*}
\includegraphics[width =\textwidth]{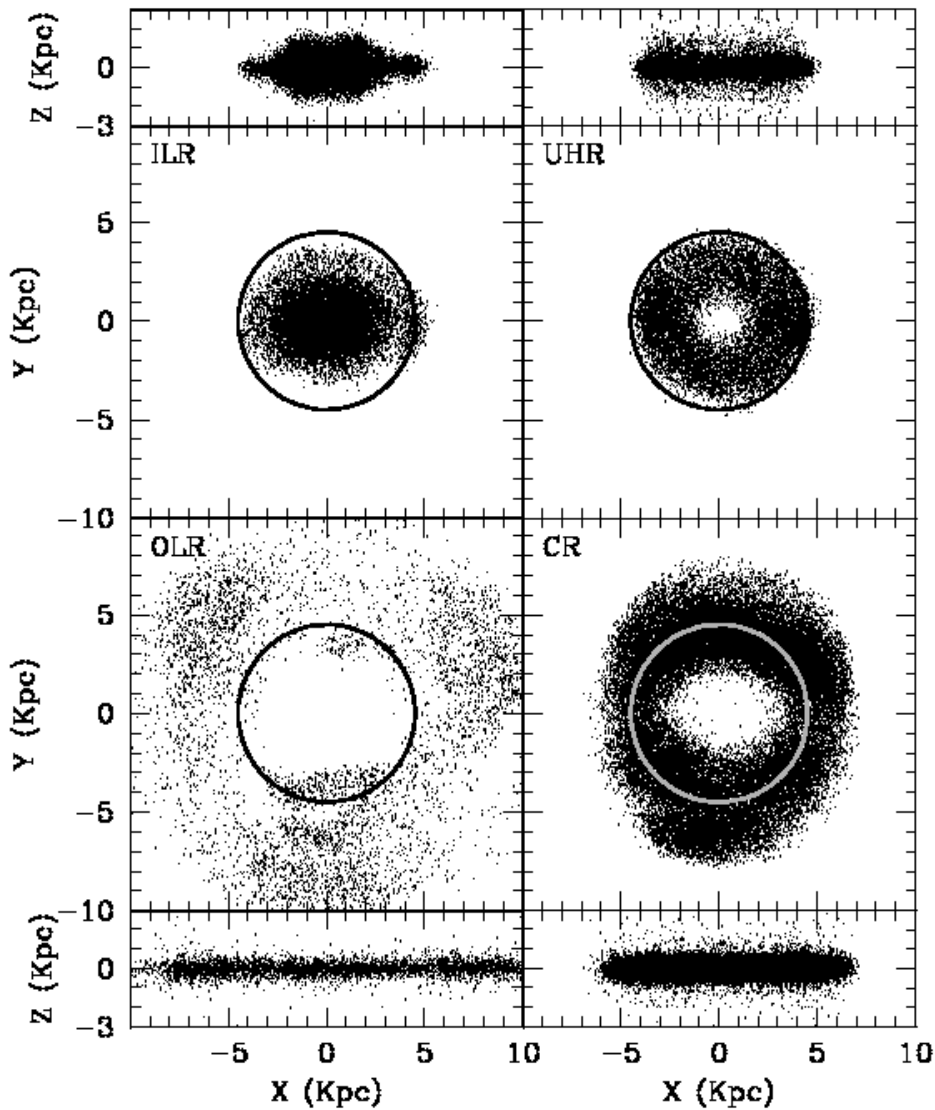}
 \caption{Spatial distributions of the particles at different
resonances in $D_{hs}$.  Each plot is labeled with the corresponding
resonance.  All resonances form rings except the ILR.  This resonance
is not localized at a given radius.  A circle marks the corotation
radius.}
\label{fig:9}
 \end{figure*} 

However, the majority of the points lies on a region spread diagonally
across the frequency map. This feature is formed by particles at
resonance, as well as particles out of resonance. It can be
approximated by the epicycle approximation up to CR.  We use the
rotation curve of each model to estimate the epicycle
frequencies.  This approximation deviates from the dynamical
frequencies inside the corotation radius.  Orbits inside the
corotation radius are elongated and their frequencies are higher than
the expected values of nearly circular orbits. This is the region
dominated by non-circular motions.

It is difficult to identify other resonances which lie on the crowded
areas of a frequency map.  As a result, we use the distribution of the
ratio $(\Omega-\Omega_{B})/ \kappa$ \citep{Atha02} to detect the
resonances of low order in l $(\pm1:m )$. Each of these resonances has
a unique value of this ratio. Fig. \ref{fig:6} shows the histogram
of $(\Omega-\Omega_{B})/ \kappa$ for the model $D_{hs}$ for a period
of 1 Gyr. The distribution has peaks at the values of different
resonances.  We found 5 of them: OLR, (1:5), CR, (-1:4) and ILR.
Fig. \ref{fig:7} shows the distribution of $(\Omega-\Omega_{B})/
\kappa$ in $K_{hb}$ for a period of 2 Gyr. As we expected from the
frequency map, the peak corresponding to the ILR is stronger than in
$D_{hs}$. This is a common feature of systems with large bars.

In these histograms, there is an underlying distribution of particles
which do not contribute to any resonant peak. They do not participate
in any resonant motion with the bar. Thus, this background of
particles should be present even before the formation of the bar. We
performed a frequency analysis of the model \ma during the first 0.5
Gyr, before the bar formation. The distribution is shown in Fig.
\ref{fig:7} as a dash curve. There are no significant peaks because
there is no bar which captures particles near resonant orbits at this
early stage of the simulation. As the simulation evolves, the bar
forms and the distribution of $(\Omega-\Omega_{B})/ \kappa$ changes
dramatically. The tail of high frequencies and high values of
$(\Omega-\Omega_{B})/ \kappa$ grows. This is produced by particles
that are sinking into the centre. However, the most dramatic change is
the formation of the resonant peaks. Some particles are captured by
the bar near some specific resonant orbits. This produces a clustering
of particles near specific dynamical frequencies and the formation of
the resonant peaks in the distribution of $(\Omega-\Omega_{B})/
\kappa$. This clustering makes that the surrounding areas outside
resonances are depopulated. This is reflected in gaps in the
distribution of $(\Omega-\Omega_{B})/ \kappa$ at both sides of a
resonance. These gaps are very clear for CR and ILR of $K_{hb}$ but
they are also visible for the (-1:5) and (1:4) resonances of $D_{hs}$.

We apply the same technique to find resonances in the halo.  We select
particles with a height from the plane lower than 3 Kpc. So, the
particles are close to the disk.  We also reject retrograde orbits.
Fig. \ref{fig:8} shows the distribution of $ ( \Omega - \Omega_{B} )
/ \kappa$ for the halo of $K_{hb}$ for a period of 2 Gyr.  The pattern
of resonances in the halo is very similar than the resonances in the
disk. In particular, CR and ILR are clearly present.

\begin{figure}
\includegraphics[width =0.55\textwidth]{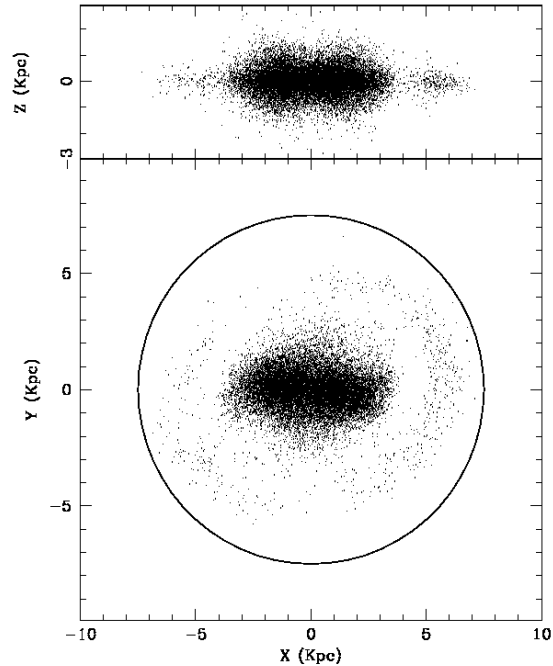}
 \caption{Spatial distribution of the particles at ILR in $K_{hb}$.
The bottom panel shows a face-on view.  The distribution resembles the
bar.  ILR particles can pass very close to the centre and span a wide
range of radii.  A circle marks the corotation radius.  the top panel
shows the vertical distribution. This edge-on view has a clear
rectangular shape.}
\label{fig:10}
\end{figure} 

\subsection{Spatial distribution of particles at resonances}

We have seen the clustering of particles near resonances in the space
of dynamical frequencies. But, what is their distribution in the
coordinate space? Are they localized around a specific radius or are
they spread over a broad region of the system? Fig. \ref{fig:9} shows
the spatial distribution of the particles at the main resonances in
$D_{hs}$.  These are the disk particles which form the main peaks in
the Fig. \ref{fig:6}. They are confined into broad areas. Thus,
particles near resonances populate a broad region of space. For
example, particles trapped at CR stay at a broad ring around the
corotation radius.  However, the distribution is not uniform along the
ring.  There are fewer particles near the ends of the bar and more
particles at both sides of the bar. This asymmetry is related to the
position of the Lagrange points of the system, as we will see in \S5.
Other rings are also formed at UHR and OLR.

However, particles at ILR do not form a ring. The ILR is not localized
around a given radius.  Fig. \ref{fig:10} shows the particles near
ILR for the model $K_{hb}$.  They are concentrated in an elongated
structure that resembles the bar. Particles at ILR can pass very close
to the centre and they have very elongated orbits. As a result, they
span a wide range of radii. This result implies that ILR extends over
a broad range of energies. This fact was outlined in different
papers. \citet{Atha03} pointed the fact that the ILR resonant orbits
are in fact the members of the x1 family of closed orbits.
\citet{Atha92} studied the energy range in which these
resonant orbits are stable so they are able to capture orbits around
them. They found that ILR orbits are stable over a significant range
of values of the Jacobi energy. \citet{WKI} also found that ILR orbits
extend up to very small radii.

In general, the vertical distribution near these resonances is very
flat.  This is expected because we are focused on planar resonances,
so we do not expect resonant motions in the vertical
direction. However, ILR is a clear exception. Its edge-on view has a
clear rectangular shape (Fig. \ref{fig:10}). This distribution again resembles
the bar. Edge-on views of N-body bars show the same rectangular shape
\citep{AthaM02}.  A frequency analysis of motions in the vertical
direction shows some particles trapped in the vertical ILR with the
same values of their radial and vertical frequencies. The trajectories
of these particles have their radial and vertical oscillations
coupled.

\section{Corotation capture}   

\begin{figure}
\includegraphics[width =0.45\textwidth]{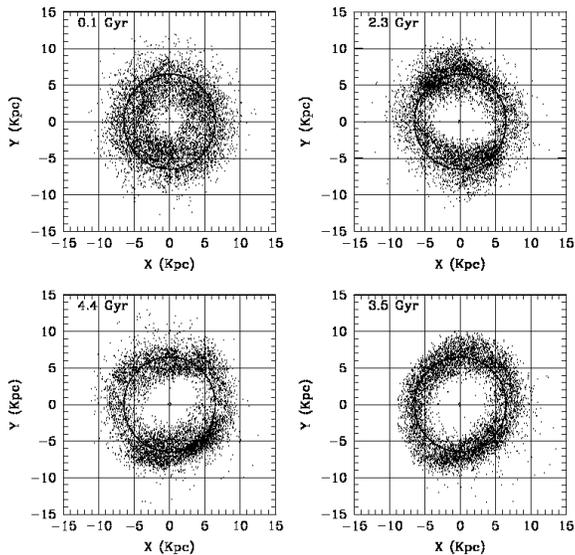}
 \caption{ Spatial distribution of the particles selected for being
near CR at 3.5 Gyr at 4 different moments. The particles are taken
from the model C.  At 0.1 Gyr, the bar is still not formed. After, bar
formation, the distribution of these particles is stationary for more
than 4 Gyr. Particles trapped near CR stay there for a long time.  The
circle represents the corotation radius.}
\label{fig:11}
\end{figure}

\begin{figure}
\includegraphics[width =0.45\textwidth]{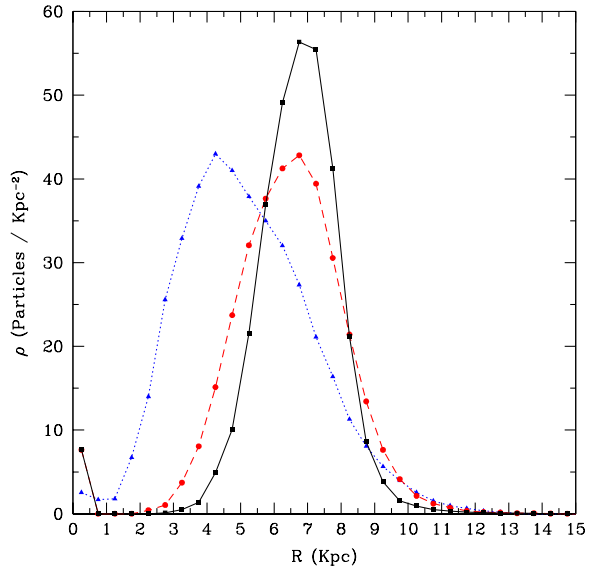}
 \caption{ Evolution of the surface density profile of particles
selected for being near CR at 3.5 Gyr for model C.  The full curve
corresponds to the distribution of the particles at 3.5 Gyr.  The dotted
curve shows the same particles at 0.1 Gyr, before the bar formation.
The dashed line is for 4.5 Gyr, 1 Gyr after the moment in which the
particles are selected for being at CR.  As the bar forms, some
particles get trapped near corotation radius. After that, the profile
shows a peak centered at the corotation radius. The distribution shows
little evolution for 4 Gyr.}
\label{fig:12}
\end{figure}

In previous sections, we saw that resonances can capture particles
with specific frequency ratios.  Now, we are going to focus on the
details of this capture mechanism and we are going to describe what
happens with the particles after being trapped near a resonance.  We
take CR, $\Omega = \Omega_{B}$, as an example, because it is easy to
visualize.  We use the model C of \citet{VK03} to study the particles
near CR.  Particles which stay near CR during a period of 1 Gyr
centred at 3.5 Gyr are selected. Fig. \ref{fig:11} and \ref{fig:12}
show their spatial distribution and their density profiles at
different moments.  At 0.1 Gyr, just before bar formation, the
particles are spread in an axisymmetric and extended distribution
around the centre. This distribution is strongly affected by the
formation of the bar. They gain angular momentum and evolve in a ring
at the corotation radius. As a result, these particles evolve when
they are being trapped at CR during the formation of the bar. After
that, the distribution is stable for more than 4 Gyr or 20 bar
rotations.  As we discussed in $4.2$, the ring is not uniform. It
shows a stronger concentration of particles in two broad areas
$90^{\circ}$ away from the major axis of the bar.  In addition, the
surface density profile of these particles shows a stationary peak
close to the corotation radius (Fig. \ref{fig:12}). These particles are still
trapped near CR several Gyr after the formation of the bar.
Therefore, CR prevents a secular evolution in the trajectories of the
particles trapped around it.

\begin{figure}
\includegraphics[width =0.45\textwidth]{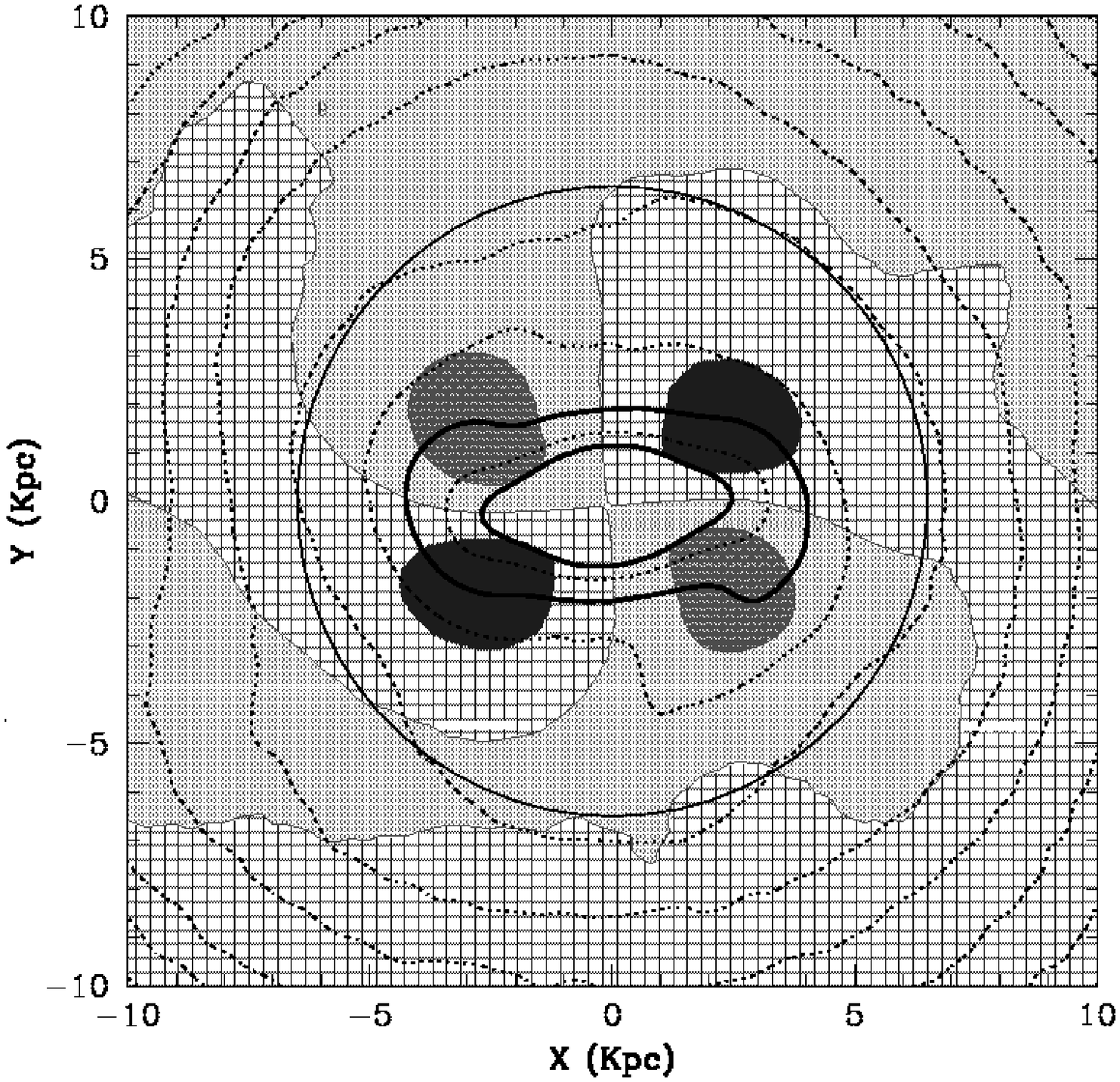}
 \caption{ Spatial distribution of the instantaneous angular momentum
change measured in a short interval of 28 Myr in model C.  See
electronic version of a color version of this plot.  The two black
leaves (dark blue) are areas with a strong negative torque.  The two
dark grey leaves (dark red) have a strong positive torque. Finally,
the broad areas filled with squares (light blue) have a small negative
torque, while the areas colored with a light grey ( light red) have a
small positive torque.  The bar is rotating in anti-clockwise
direction.  The particles which are close to the bar and moving ahead
of the bar, loose angular momentum (black/blue areas).  At the same
time, particles which are moving behind the bar, gain angular momentum
(dark grey/red areas).}
\label{fig:13}
 \includegraphics[width =0.45\textwidth]{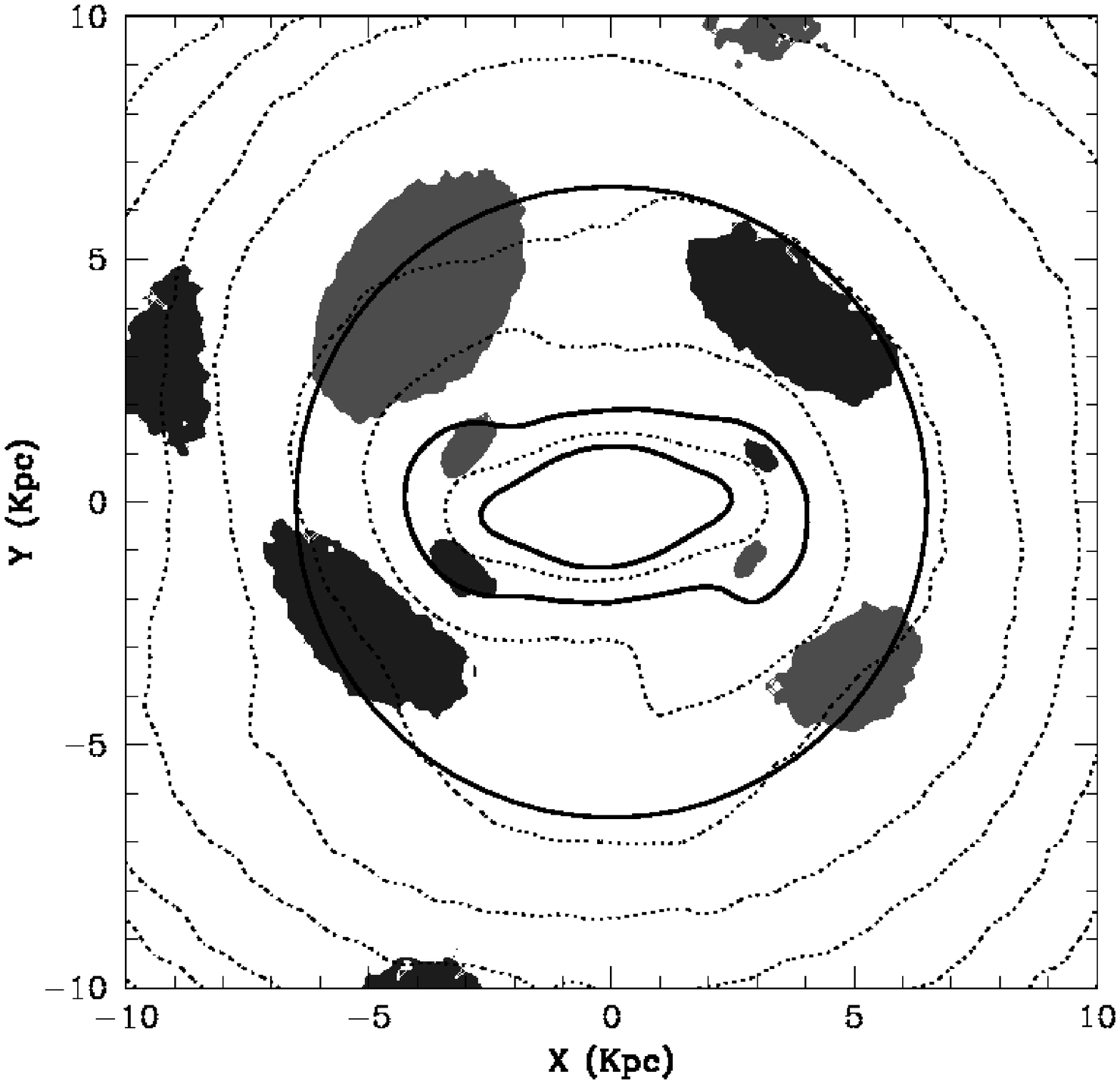}
 \caption{ Distribution of the peaks in the change in angular momentum
in the disk for a longer period of time (150 Myr).  The grey (red)
areas show the position of the particles with the maximum increment in
their angular momentum.  The black (blue) areas show the position of
the particles with the maximum decrement of angular momentum.  Those
areas lie mainly along the corotation radius (black circle)}
\label{fig:14}
\end{figure}

We can now study in detail how CR prevents the secular evolution of
trapped particles in an almost stationary model, in which $\Omega_B$
is almost constant. We start with the spatial distribution of the
change in angular momentum of particles in the disk. We compute the
change of angular momentum of each particle between two different
moments.  That change contributes to the average angular momentum
change at the position of the particle in an intermediate moment. The
result is a field of angular momentum change.  If these two moments
are very close, that field is a good approximation of the
instantaneous change of angular momentum in the disk. In other words,
we are measuring the torque field produced by the bar and the spiral
arms (Fig. \ref{fig:13}). This torque has a similar shape of a
torque from an elongated structure in rotation. It shows a
$180^{\circ}$ symmetry. The bar is rotating in anti-clockwise
direction.  So, the particles which are close to the bar and moving
ahead of the bar, are being pushed backwards by the bar. So, they
loose angular momentum (black areas in Fig. \ref{fig:13}). At the same time,
particles which are moving behind the bar, are being pushed
forwards. So, they gain angular momentum (dark grey areas in Fig.
\ref{fig:13}).


\begin{figure}
 \includegraphics[width =0.45\textwidth]{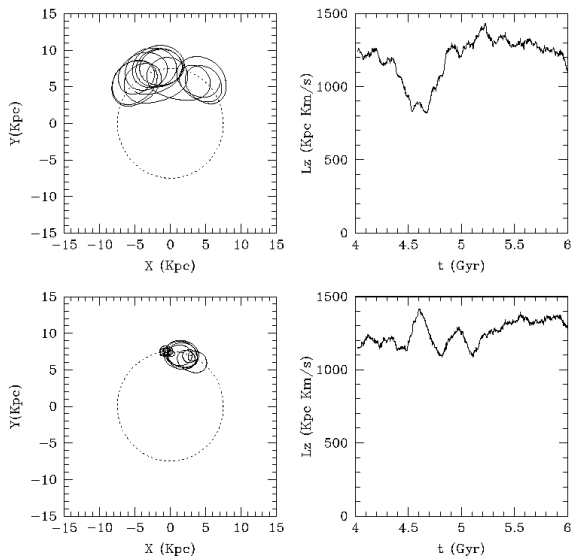}
 \caption{ Example of two trajectories of particles near CR for 2 Gyr
in the model $K_{hb}$.  They are selected for being near CR between 4
Gyr and 4.5 Gyr.  The left panels show their trajectories in the
reference frame which rotates with the bar.  The bar is always along
the horizontal axis. The dotted circle is at the corotation radius.
The example at the top is a trajectory in libration around an stable
Lagrange point The trajectory at the bottom is even closer to this
point.  The right panels show the angular momentum of both
trajectories over 2 Gyr.  The angular momentum oscillates, but there
is no overall change of angular momentum.  These trajectories do not
evolve.  As a result, CR prevents the evolution for orbits trapped
around it.}
\label{fig:15}
\end{figure}

However, small changes in angular momentum over short periods of time
can accumulate over a longer period. This can produce a much stronger
change in angular momentum over a long period of time.  Fig.
\ref{fig:14} shows only the areas with the largest variations in
angular momentum over a much longer interval of time than in Fig.
\ref{fig:13}. The selected interval is nearly one period of the bar,
150 Myr.  The choice of different intervals produces similar results.
The largest change of angular momentum lays mainly near the corotation
radius and away from the major and minor axes of the bar. The two
areas of positive change of angular momentum do not cover the same
area. One is bigger than the other. However, this asymmetry is not
permanent. It oscillates slowly with time. As a result, the net effect
of this asymmetry is averaged out.  Other peaks are also found inside
the bar, a region dominated by ILR orbits. Finally, other peaks beyond
the corotation radius could be attributed to the OLR.

Near the corotation radius, particles which are moving behind the bar
for a significant interval of time accumulate small increments of
angular momentum over that interval. The net effect is a significant
increase in their angular momentum. The opposite is true for particles
which stay ahead of the bar for a long period of time. They loose a
significant amount of angular momentum. The main question is now how
this change in angular momentum affects the evolution of these
particles. The na\"{i}ve idea is that a particle that increase its angular
momentum evolves strongly. However, a particle initially in one of the
areas of maximum positive change in angular momentum near CR (grey/red
areas in Fig. \ref{fig:13}) gains angular momentum and moves
outwards. As a result, the particle rotates slower than the bar, $
\Omega<\Omega_B$. It lags the bar and starts to move to the area of
negative angular momentum change (black/blue areas in Fig.
\ref{fig:13}). In that area, the particle looses the angular momentum
gained previously and moves inwards. As a result, it starts to rotate
faster than the bar, $ \Omega>\Omega_B$, and to move towards the area
of positive change of angular momentum.  In that area, the particle
gains angular momentum and the cycle starts again. As a result,
the changes in angular momentum are compensated along the trajectory
of each particle near CR. These oscillations of angular momentum
produces no net torque over a long period of time. CR actually
prevents the evolution of the particles trapped near the corotation
radius.

Fig. \ref{fig:15} provides two examples which illustrate this
previous idea. They are taken from the model $K_{hb}$. The particles
are selected to be near CR between 4 and 4.5 Gyr. Then, we plot their
trajectories for the next 2 Gyr. We select the non-inertial frame
which rotates with the bar. In that frame, the bar is always along the
horizontal axis and a particle exactly at CR is not moving in this
frame. The top panels of Fig. \ref{fig:15} show a particle which
slowly oscillates once along a section of the corotation ring during 2
Gyr.  The particle moves between the areas in which the change of
angular momentum is positive and negative. As a result, the changes in
its angular momentum cancel each other. The net result of this
oscillation is an almost no change in the angular momentum. The
particle is trapped in an orbit of libration for several Gyr and many
orbits.

The bottom panels of Fig. \ref{fig:15} show an orbit closer to
CR. It also librates at the corotation radius but the amplitude of the
oscillations are much smaller than in the previous case. Actually,
during the last Gyr of the simulation, the particle spends several
orbits around a single point in the rotating frame. The angular
momentum during this time is almost constant. So, the net transfer of
angular momentum near that point is minimum. In other examples, the
orbits librate in a similar way but around another point on the other
side of the corotation ring.

This trapping mechanism near CR is actually well known in galactic and
celestial mechanics.  The situation is equivalent to the
motions of arbitrary amplitude around the stable Lagrange points of a
stationary non-axisymmetric system. The equations of motion can be
reduced to the equations of a pendulum \citep[Chapter 3,
eqs. (3-123)-(3-129) ]{BT87}. Particles trapped near CR librate slowly around
one of the stable Lagrange points. A particle exactly at this point is
exactly at CR. A close example of such resonant orbit is shown in the
bottom panels of Fig.  \ref{fig:15}. This orbit moves along the
corotation radius with the same angular velocity of the bar
rotation. CR keeps particles in libration orbits for a long time and
for many periods. While particles are trapped, their orbits do not
evolve. As a result, CR prevents the evolution of the particles
trapped around it and minimize their angular momentum transfer.


\section{Comparison with an analytical model}

\begin{figure}
 \includegraphics[width =0.45\textwidth]{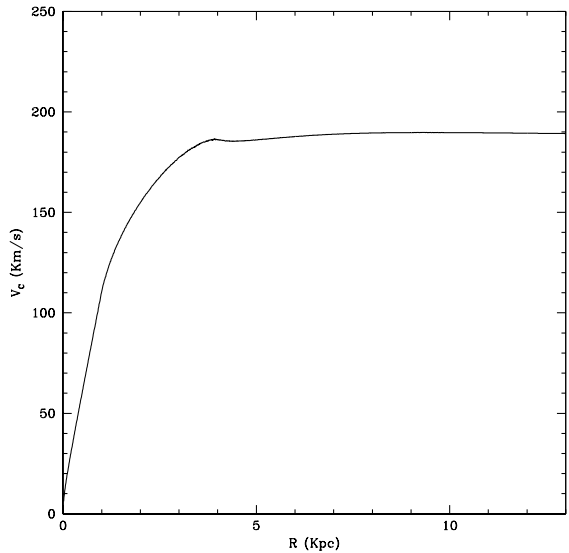}
 \caption{ Circular velocity profile of the analytical model of a
barred galaxy.  The model corresponds to a typical high surface
brightness barred galaxy with a maximum circular velocity close to 190
km s$^{-1}$.}
\label{fig:16}
\end{figure}

As we saw in the previous section, CR captures particles around two
stable Lagrange points.  The stability around these points is mainly
determined by the topology of the gravitational potential around
them. Therefore, we can approximate this potential with a simple
analytical model which captures the main features of a strongly barred
system. Such a model can be used to clarify the results of N-body
simulations. The analytical potential consists of a Miyamoto disk, a
NFW halo and a homogeneous prolate ellipsoid. The ellipsoid rotates
with a given pattern speed $\Omega_B.$ The corresponding expressions
are described in appendix A and the parameters of the model are given
in Table 3. It reproduces a strongly barred galaxy. Its circular
velocity profile is shown in Fig. \ref{fig:16}. Using this
background potential, we can compute the trajectories of a set of
particles and follow their resonant interactions with the bar. This
analytical model can catch some fundamental aspects of resonances,
like trapped orbits near resonances. At the same time, this simple
model does not have the inherent complexity of a self-consistent
N-body simulation. As a result, it is easy to interpret. Then, this
interpretation can be used to understand better the resonant phenomena
in more complex and realistic cases, like in N-body models.
\begin{table} 
\caption{Set of parameters for the analytical model.}
 \begin{center}       
 \begin{tabular}{|lllcccc} \hline     
\multicolumn{2}{l} {Miyamoto disk:} \\\hline Disk Mass ($10^{10}
 \Msun$) & 5 \\ A (kpc) & 4 \\ B (kpc) & 1 \\ \hline NFW halo: \\
 \hline $r_s$ (Kpc) & 28.7 \\ $\rho_0$ ($10^{6} \Msun $ Kpc$^-3$ ) &
 3.12 \\ \hline Prolate ellipsoid: \\ \hline Bar mass ($10^{10}
 \Msun$) & 1 \\ Semimajor axis (Kpc) & 4 \\ Semiminor axis (Kpc) & 1
 \\ Pattern speed (km s$^{-1}$ Kpc$^{-1}$) & 36 \\
 \end{tabular} 
 \end{center}
 \end{table}
\begin{figure}
\includegraphics[width =0.45\textwidth]{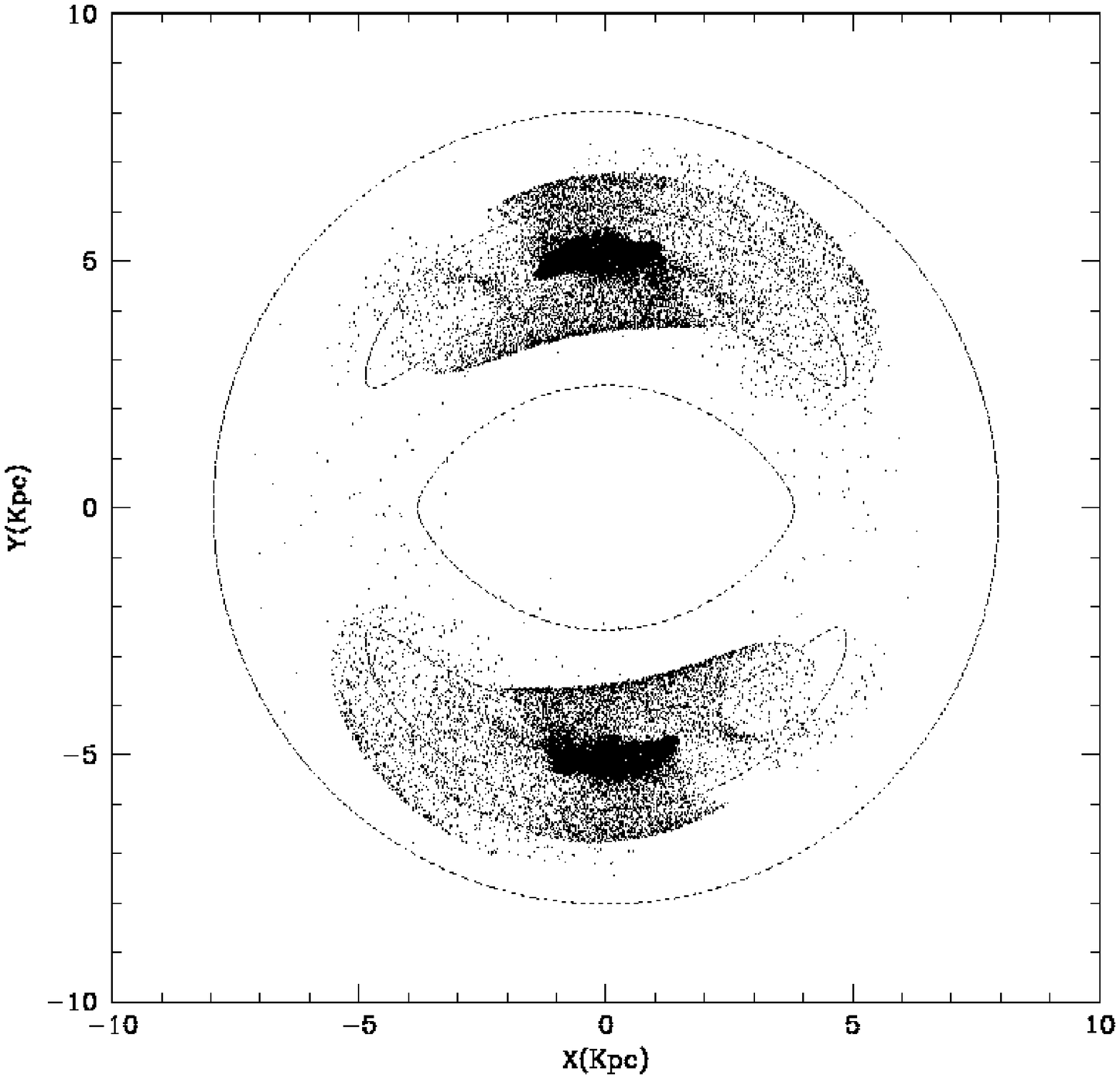}
\caption{ Spatial distribution of the particles trapped at CR after
100 bar rotations (17 Gyr) The distribution fills two lobes around the
stable Lagrange points.  It resembles the distribution of particles
trapped near CR in the N-body models (Fig. \ref{fig:9} and
\ref{fig:11}).  The background contours are contours of equal
effective potential.}
\label{fig:17}
\includegraphics[width =0.45\textwidth]{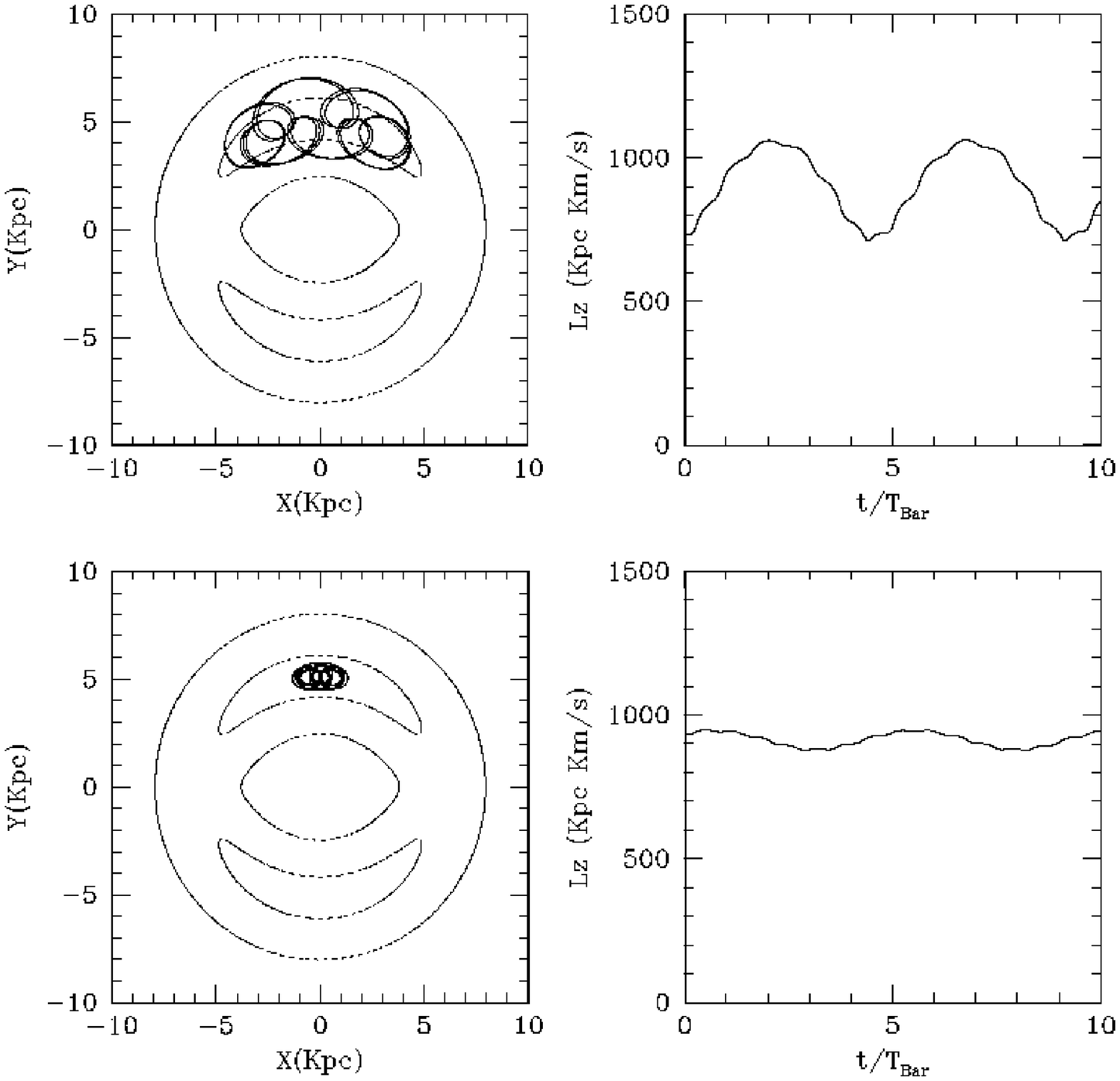}
\caption{Two examples of trajectories trapped near CR, as in Fig.
\ref{fig:15}.  The trajectories librate around the stable Lagrange
point and their angular momentum oscillate with no net change of
angular momentum after approximately 5 bar periods.}
\label{fig:18}
\end{figure}


Fig. \ref{fig:17} shows particles trapped near CR at two stable
Lagrange points. The initial configuration is a set of $32^3$
particles distributed uniformly inside two small spheres of 0.1 Kpc in
radius. Each one is centered at each stable Lagrange point. The
distribution of velocities is initially centered on the circular
velocity at the corotation radius. The dispersion is 25 per cent of that
velocity. After 100 bar rotations, almost all particles fill two
banana-shape areas at both sides of the bar.  They are still trapped
around the stable Lagrange points.  They cover a broad area along
the corotation radius. This supports the results discussed in $\S5$
for a self-consistent N-body experiment. The distribution of the
particles near CR in Fig. \ref{fig:9} and \ref{fig:11} resembles the
distribution of the particles trapped around the two stable Lagrange
points in Fig. \ref{fig:17}.  The region of trapped particles is not
an infinitesimal volume around the stable points. It extends over a
broad volume both in spatial and phase space. The distribution of the
energies of these particles cover a range which is 10 per cent of the
energy at the Lagrange point.  CR covers a significant volume in
phase space, although the only particle formally at resonance is the
one which moves with one of the stable points.

In general, these trapped particles move in libration orbits around
the Lagrange points. They are trapped for a cosmologically
significant period of time. Fig. \ref{fig:18} shows two examples of
libration orbits and the change of their angular momentum. These
examples can be compared with the examples taken from the N-body model
and discussed in \S5.  The first example has large amplitude of
libration. As a result, the angular momentum oscillates
significantly. However, the angular momentum oscillates and the net
change is canceled in a libration period. As a result, the net
angular momentum transfer over many orbits is zero.  The bottom panels
of Fig. \ref{fig:18} show another example. Its amplitude of
libration is much smaller than in the first example.  The oscillations
in the angular momentum are also much smaller.  This example is
similar to the second example of \S5.  The particle is closer to CR
and remains in a stable orbit which minimizes the transfer of angular
momentum. This supports the interpretation that resonances prevent the
orbital evolution of trapped particles.

\begin{figure}
\includegraphics[width =0.45\textwidth]{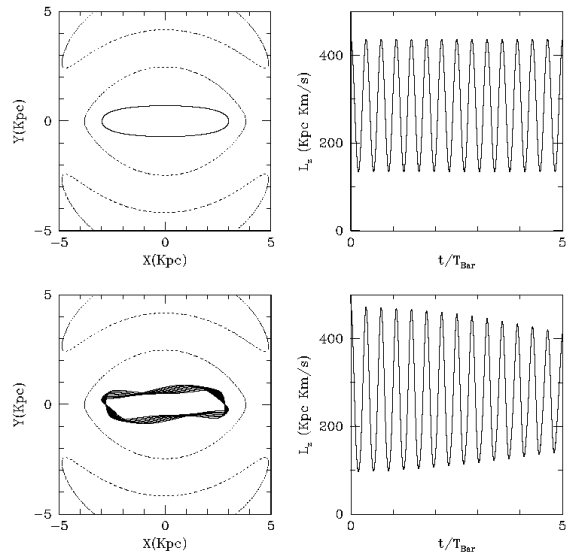}
\caption{Two examples of trajectories as in Fig. \ref{fig:15} but
for the ILR.  The top panels show an example of a ILR orbit.  The
angular momentum oscillates but average change of angular momentum is
zero.  The bottom panels show one example of a trajectory trapped
around the ILR orbit.  The trajectory librates around the closed
orbit.  The angular momentum is modulated by this libration but again
the net average change over many periods is zero.}
\label{fig:19}
\end{figure}

Similar conclusions can be obtained for other resonances.  In the case
of the ILR, the resonant orbits are closed orbits elongated along the
bar.  The top panels of Fig. \ref{fig:19} show an example of a ILR
orbit.  The angular momentum oscillates significantly but the net
change of angular momentum averaged over one orbit is zero.  As we
pointed in $\S4.2$, the ILR orbits are known as the $x1$ family. They
are the backbone of the orbits which support the bar
\citep{Skokos,Atha92}. These resonant orbits capture particles around
them in a similar way corotation does. The bottom panels of Fig.
\ref{fig:19} show one example of a trajectory trapped around the ILR
orbit. The trajectory librates around the closed orbit.  The angular
momentum is modulated by this libration but again the net average
change over many periods is zero.  The main difference between CR and
ILR is that ILR orbits are a set of orbits instead of two Lagrange
points. The result is that the ILR resonance covers a larger volume in
phase space which cover the bar, as we saw in the distribution of ILR
particles in Fig. \ref{fig:10}. The trajectories in Fig. \ref{fig:19} are
stable forever. We tracked them for more than 100 bar rotations and
there is no indication of evolution in their trajectories. As we saw
before, particles trapped at resonance do not evolve. For the ILR,
particles are trapped by a set of $x1$ orbits and forced to move
within the bar as the bar rotates.

\begin{figure}
\includegraphics[width =0.45\textwidth]{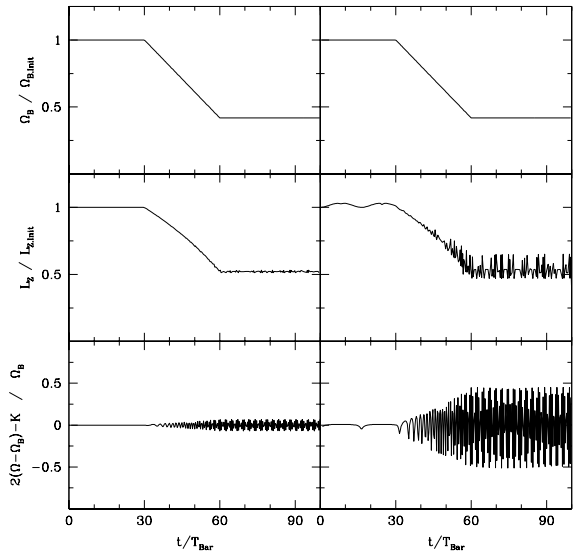}
\caption{Left and right panels show two different examples of
particles near ILR when the bar slows its rotation during 30 bar
rotations between 30 and 60 bar rotations.  Top panels show the
evolution of the bar pattern speed.  Middle panels show the evolution
of the angular momentum averaged over a single orbit. Bottom panels
show the ILR condition (eq. 2) scaled by $\Omega_B$.  During the
slowdown of the bar, the trajectory exactly at ILR(left) and the
trajectory trapped near ILR(right) remain near resonance. They track
the resonance and follow the bar evolution.}
\label{fig:20}
\end{figure}

However, what happens with the particles trapped at resonance when the
bar evolves? In this case, the resonant structure also evolves. The
volume in the phase-space which satisfies a particular resonant
condition drifts through the phase-space. Does the trajectory
initially trapped at a resonance follow it or is scattered off
resonance? In order to address this issue, we follow the same ILR
trajectories when the bar slows its rotation. The top panels of Fig.
\ref{fig:20} show the imposed bar pattern speed. It remains constant
for 30 initial bar rotations (5 Gyr), then it decreases linearly for
other 30 initial bar rotations and finally remains constant for 40
initial bar rotations. At the end, the bar pattern speed decreases to
40 per cent of its original value over 5 Gyr. From each trajectory, we
extract the specific angular momentum averaged over time for an orbit
in the rotating frame and its radial and angular frequencies over the
orbit. The radial frequency is computed as $ \kappa = 2 \pi / T_p $,
where $T_p$ is the period of the radial oscillations, defined as the
time between apocenters. At the same time, the angular frequency is
defined as $ \Omega = \Delta \Phi / T_p $, where $\Delta \Phi$ is the
angle swept by the trajectory between apocenters in the inertial
frame. Using these frequencies and the pattern speed of the bar,
we can check the resonant condition for ILR (Eq. 2).  

The left side of
Fig. \ref{fig:20} shows the orbit which is initially trapped exactly
at ILR.  Before any evolution, the averaged angular momentum remains
constant and the resonant condition is fulfilled. During the slowdown
of the bar, the angular momentum of the trajectory decreases at almost
the same rate of the change in the pattern speed. The trapped particle
follows the slowdown of the bar. More important is the fact that the
trajectory remains close to the resonance although the system
evolves. The particle is trapped at resonance and oscillates around
the resonant orbit as the system evolves as a whole. This tracking of
a resonance is true even for trajectories further from the exact
resonant orbit but still trapped around it. The right side of Fig.
\ref{fig:20} shows an example. The oscillations around the resonant
condition are larger than in the previous case, but the trajectory is
not scattered off resonance when the system evolves. The same results
can be found for bigger rates of change in the pattern speed. Only
when the pattern speed decreases to one half of its original value in
less than one bar period, the oscillations around the resonant
condition become wider.

\section{Discussion}

Resonances play an important role in barred galaxies. Stable resonant orbits
can capture disk and halo material in near-resonant orbits. The bar
itself is a manifestation of this resonant capture. The $x1$ family of
closed orbits are ILR resonant orbits that capture particles in
elongated orbits along the bar major axis \citep{Atha03}. These orbits
support the orbital structure of the bar \citep{Skokos,Atha92}. 

Resonances prevent the dynamical evolution of
the material trapped near resonant orbits. This material does not
experience a net change of angular momentum although the angular
momentum oscillates strongly over an orbital period (Fig.
\ref{fig:19}). A particle exactly at resonance with the bar has
no evolution in its trajectory. The change in angular momentum over an
orbital period is zero. The particle exactly at resonance stay at a
resonant closed orbit forever. As a result, resonances tend to
minimize the exchange of angular momentum between trapped material and
the bar.

Therefore, the mere presence of resonances in barred galaxies do not
drive their secular evolution. Orbits trapped at resonance only evolve
if the bar evolves as a whole (Fig. \ref{fig:20}). In this case,
resonances drift as the bar evolves. Particles anchored near resonant
orbits track the resonances and consequently evolve. As a result, the
evolution at resonances is linked to the evolution of the bar. For
example, if the bar slows its rotation, the Lagrange points move
outwards. As a result, CR moves outwards and trapped particles, which
track the motion of CR, move also outwards. The result is that these
particles near CR gain angular momentum and evolve. At the same time,
ILR particles trapped by the bar loose angular momentum. This angular
momentum is not lost because the particles are near a resonance. It is
lost because there is a net torque, which produces a slowdown of the
bar and of the trapped particles trapped near ILR. 

This torque can be the
result of the dynamical friction with the dark matter halo. Resonances
in the halo can also play an important role in this interaction. 
ILR particles in the halo form an ellipsoidal structure in 
the inner halo.
\citep{CVK06}. This halo bar exerts a torque on the stellar bar
that slows its rotation. This is an interaction between different
structures trapped at ILR. As a result of this interaction, ILR
particles in the stellar bar can loose angular momentum and ILR
particles in the halo can gain it. This interpretation is consistent
with the results on the angular momentum change near resonances in
the case of the bar slowdown \citep{Atha03}. However, this
mechanism of change of angular momentum at resonances is very
different from the resonant transfer of angular momentum predicted in
perturbation theory \citep{Atha03,LBK72}.

The resonances found in \citet{Atha03} are consistent with the
resonances found in our models. However, the models in \citet{Atha03}
have a stronger ILR, CR was weaker and outer resonances like OLR were
almost absent. These differences are due to differences in the
corotation radius. Based on the initial rotation curve and the pattern
speed of the model MQ2 of \citet{Atha03}, the corotation radius was
roughly 20 Kpc, 5.7 times its disc scalelength. This is 5 times larger
than the corotation radius in our model $D_{hs}$.  Thus, the models in
\citet{Atha03} have most of the material of the disk within the
corotation radius. This material can only be captured by inner
resonances like ILR. That results in a strong ILR peak. On the other
hand, little material lies at corotation radius and beyond. Therefore,
CR, OLR and other outer resonances can trap only a few particles. This
is why their peaks in the $(\Omega-\Omega_{B})/ \kappa$ histogram of
\citet{Atha03} are much smaller than in our model $D_{hs}$.

Recently, \citet{WKI} discussed the dynamics of the interaction
between a bar and a dark matter halo and described the requisites
needed to follow this resonant interaction accurately. The first
criterion stated that a N-body model should have enough phase-space
coverage at resonances. In order to ensure the correct resonant
behavior, the phase space near resonances should be populated with
enough particles. The phase-space volume near resonances is defined by
the separatrix which divides trapped from non-trapped orbits in the
phase space near a resonance. This volume covers the region of trapped
orbits which librate around the stable resonant orbit. \citet{WKI}
argue that 10 particles inside one tenth of that resonant region are
enough to obtain the correct behavior at resonances. Our model
$D_{hs}$ has around $1.5 \times 10^4$ disk particles near ILR and $4.1 \times
10^4$ near CR. These are trapped particles which stay in the
libration region near resonances. As a result, resonances are well
populated in our N-body models. This is because the region of trapped
orbits is large (Fig. \ref{fig:6}, \ref{fig:7}, \ref{fig:8}).

The second criterion deals with the artificial noise in the potential
and the effects of two-body scattering. They can introduce an
artificial diffusion of orbits. The characteristic diffusion length
should be smaller than the resonance width, defined by the size of the
region of trapped orbits. Otherwise, particles can artificially
diffuse out of a resonance. We have seen that these regions near
resonances are large. So, this second criterion is also achieved.

\section{Summary and conclusions}

We have detected dynamical resonances in N-body models of barred
galaxies with evolving disks in live dark matter halos. The dominant
resonances are the corotation resonance (CR) and the inner Lindblad
resonance (ILR), although other low order resonances, like the outer
Lindblad resonance (OLR) or the ultra-harmonic resonance (UHR) are
also present (Fig. \ref{fig:6}, \ref{fig:7}, \ref{fig:8}). Resonances
in the halo are also found.

In general, resonances cover broad areas of the disk (Fig.
\ref{fig:9}). Particles at CR are distributed in a wide ring at the corotation
radius. On the other hand, the epicycle frequencies are not equal to
the natural frequencies of particles inside the corotation radius,
where non-circular motions are important. This is specially true for
the ILR. this resonance is not localized at a given radius. Particles at ILR
are found mainly in elongated orbits inside the bar. Their
distribution resembles the bar (Fig. \ref{fig:10})

In all three studied models, we find that resonances
capture particles and force them to move in trajectories near stable
resonant orbits. For example, the corotation resonance traps particles
in libration orbits around the two stable Lagrange points of the
system. As a result, the angular momentum oscillates with the period
of the libration motion but the net change over many orbits is
zero. Therefore, these trapped particles do not evolve. We conclude
that resonant trapping tends to minimize the change of angular
momentum of the particles trapped around them.  However, the trapped
particles can participate in the global evolution of the bar because
they are locked at resonances. They are still trapped during the
slowdown of the bar (Fig. \ref{fig:20}). As a result, ILR particles
can loose angular momentum during this slowdown. Particles trapped at
resonances only evolve when the bar evolves as a whole.

\section*{Acknowledgments}

We thanks O. Valenzuela and M. Weinberg for many useful discussions.
We acknowledge support from the grant NSF AST-0407072 to NMSU.
The computer simulations presented in this paper were performed at the
National Energy Research Scientific Computing Center (NERSC) of the
Lawrence Berkeley National Laboratory and the NASA Advanced
Supercomputing (NAS) Division of NASA Ames Research Center.

\bibliographystyle{mn2e}
\bibliography{reso}

\appendix

\section{An analytical galactic potential}

The 3D potential used in \S6 consists of three components. First, A
Miyamoto potential represents the disk:
\begin{equation}
\Phi_D(\mathbf{x}) = - \frac{GM_D}{\sqrt{ x^2 + y^2 + ( A +
\sqrt{B^2+z^2})^2}}
\end{equation}
where $M_D$ is the mass of the disk, A and B are the horizontal and
vertical scalelengths. The dark matter halo is modeled using a NFW
halo:
\begin{equation}
\Phi_H(\mathbf{x}) = -4 \pi G \rho_s r_s^2 \frac{ \ln(1+r/r_s)}{r/r_s} ,
\ r=\sqrt{x^2+y^2+z^2}
\end{equation}
Finally, the contribution of the bar is modeled as a prolate
homogeneous ellipsoid. The potential of a point inside the ellipsoid
is given by
\begin{equation}
 \Phi_B(\mathbf{x}) = -\pi G \rho \left( I(\mathbf{a}) a_1^2 - \sum
 A_i(\mathbf{a}) x_i^2 \right)
\end{equation}
where $x_i=\{x,y,z\} , i=1,3 $. $ a_1 $ is the semimajor axis of the
bar and $a_2=a_3$ is the semiminor axis.  $ I(\mathbf{a}) $ and
$A_i(\mathbf{a}) , i=1,3$ are defined by the relations:
\begin{equation}
I(\mathbf{a}) = \frac{1}{e} \ln \frac{1+e}{1-e}
\end{equation}
\begin{equation}
e=\sqrt{1 - \frac{ a_1^2}{a_3^2} } 
\end{equation}
\begin{equation}
A_1(\mathbf{a}) = \frac{1-e^2}{e^2} \left[ \frac{1}{1-e^2} - \frac{1}{2e}
\ln \frac{1+e}{1-e} \right]
\end{equation}
\begin{equation}
A_2(\mathbf{a}) = A_1(\mathbf{a})
\end{equation}
\begin{equation}
A_3(\mathbf{a}) = 2 \frac {1-e^2}{e^2} \left[ \frac{1}{2e} \ln
\frac{1+e}{1-e} -1 \right]
\end{equation}
Similarly, the potential of a point outside the ellipsoid is given by:
\begin{equation}
 \Phi_B(\mathbf{x}) = -\pi G \rho \frac{a_1a_2a_3} {\acute{a}_1
 \acute{a}_2 \acute{a}_3 } \left( I(\mathbf{\acute{a}}) \acute{a}_1^2 -
 \sum A_i(\mathbf{\acute{a}}) x_i^2 \right)
\end{equation}
where 
\begin{equation}
  \acute{a}_i^2 = a_i^2 + \lambda(\mathbf{x}) : \sum \frac{x_i}{
  \acute{a}_i^2 } = 1.
\end{equation}

\bsp

\label{lastpage}

\end{document}